\documentclass[a4paper,11pt]{article}
\usepackage{jcappub} 
\usepackage{cancel}
\usepackage{siunitx}
\usepackage{orcidlink}
\usepackage{comment}
\usepackage[normalem]{ulem}
\usepackage{booktabs}
\usepackage{appendix}
\usepackage{indentfirst}
\usepackage{xcolor}
\usepackage{amsmath,mathtools}
\usepackage[T1]{fontenc} 
\usepackage{hyperref}
\usepackage{tikz}
\usepackage{booktabs}

\title{A spherical hydrodynamical model of cosmic voids in $\Lambda$CDM and beyond}

\author[a,b]{Tommaso Moretti\,\orcidlink{0009-0006-4815-4764},}
\author[c]{Giovanni Verza\,\orcidlink{0000-0002-1886-8348},}
\author[a,b]{Noemi Frusciante\, \orcidlink{0000-0002-7375-1230},}
\author[d,e,f]{Francesco Pace\, \orcidlink{0000-0001-8039-0480}}

\affiliation[a]{Dipartimento di Fisica “E. Pancini”, Università degli Studi di Napoli “Federico II”,\\Compl. Univ. di Monte S. Angelo, Edificio G, Via Cinthia, I-80126, Napoli, Italy}

\affiliation[b]{INFN Sezione di Napoli, Università degli Studi di Napoli “Federico II”,\\Compl. Univ. di Monte S. Angelo, Edificio G, Via Cinthia, I-80126, Napoli, Italy}

\affiliation[c]{Center for Computational Astrophysics, Flatiron Institute, 162 5th Avenue, 10010, New York, NY, USA}

\affiliation[d]{Dipartimento di Fisica, Università degli Studi di Torino, Via P. Giuria 1, I-10125 Torino, Italy}

\affiliation[e]{INFN-Sezione di Torino, Via P. Giuria 1, I-10125 Torino, Italy}

\affiliation[f]{INAF-Istituto Nazionale di Astrofisica, Osservatorio Astrofisico di Torino,
strada Osservatorio 20, 10025, Pino Torinese, Italy}

\emailAdd{tommaso.moretti@unina.it}
\emailAdd{gverza@flatironinstitute.org}
\emailAdd{noemi.frusciante@unina.it}
\emailAdd{francesco.pace@unito.it}

\abstract{
Cosmic voids have emerged as powerful probes for cosmology, providing complementary information on the large-scale structure of the universe.  
We present the first application of a hydrodynamical framework to model the evolution of cosmic voids. This approach offers a physically intuitive characterization of void dynamics and can naturally be applied to non-standard cosmologies. 
We derive the cosmology-dependent mapping that relates the linear (Lagrangian) and fully non-linear (Eulerian) evolution of the matter density contrast, a central component for accurate theoretical modeling of void statistics. Furthermore, we present a new method for determining the shell-crossing epoch across arbitrary cosmological backgrounds, thereby extending previous treatments restricted to the Einstein–de Sitter universe. 

Motivated by recent DESI results hinting at dynamical dark energy, we investigate void evolution in $ w_0w_a$CDM cosmologies by varying $ w_0$ and $w_a$. We also consider the impact of varying the matter density parameter, $ \Omega_{\mathrm{m},0}$. We find that the evolution of isolated, spherically symmetric cosmic voids is most sensitive to $ \Omega_{\mathrm{m},0} $ and $ w_0 $, which can alter the non-linear density contrast by up to 20--30\%. Variations in $w_a$ have a smaller impact, but may still lead to measurable effects. We also show that the cosmology-dependent mapping between linear and non-linear density contrasts may provide a sensitive probe of dynamical dark energy in precision void analyses.
}
\begin{document}
\maketitle
\flushbottom

\section{Introduction}
Cosmic voids are vast, underdense regions in the large-scale structure (LSS) of the universe and were first identified in galaxy redshift surveys in the late 1970s~\cite{Gregory:1978qwe,Joeveer:1978ers}. Since then, they have  been recognized as fundamental components of the Cosmic Web~\cite{Zeldovich:1982zz,Bond:1995yt}. Typically spanning radii of $(20$--$50)\,h^{-1}\mathrm{Mpc}$, voids occupy about 80\% of the observable volume of the universe~\cite{vandeWeygaert:2014mqv,Cautun:2014fwa}, yet contain only a small fraction of its dark matter, diffuse baryonic gas, and galaxies~\cite{Chincarini_1975,Einasto:1980qwe,Kirshner:1981qwe,Colless:2003wz,SDSS:2003tbn,Guzzo:2013spa}. With the advent of large and deep redshift surveys~\cite{EUCLID:2011zbd,DESI:2016fyo,LSSTScience:2009jmu}, voids have emerged as powerful cosmological probes~\cite{Pan:2012xxx,Sutter:2012rts,Brouwer:2018xnj,Euclid:2023eom,Pisani:2019cvo,Moresco:2022phi,Fraser:2024ecp,Contarini:2022nvd,Woodfinden:2022bhx,Hamaus:2016wka,Mao:2016faj,Achitouv:2016mbn,Hawken:2016qcy,Hamaus:2017dwj,Achitouv:2019xto,Nadathur:2019mct,Hamaus:2020cbu,eBOSS:2020nuf,Hawken:2019rpp,eBOSS:2020yxq,Woodfinden:2023oca,Contarini:2022mtu,Song:2025vjz}, capable of constraining a wide range of physical phenomena including dark energy (DE)~\cite{Lee:2007kq,Lavaux:2009wm,Bos:2012wq,Pisani:2015jha,Verza:2019tvg,Biswas:2010ey}, modified gravity (MG)~\cite{Perico:2019obq,Clampitt:2013tyg,Li:2011qda,Spolyar:2013maa,Pollina:2015uaa,Zivick:2014uva,Achitouv:2016jjj,Sahlen:2015wpc,Barreira:2015vra,Cai:2014fma,Voivodic:2016kog,Davies:2019yif,Paillas:2018wxs,Contarini:2020fdu,Mauland:2023eax,Wilson:2022ets,Maggiore:2025mbp}, neutrinos~\cite{Massara:2015msa,Banerjee:2016zaa,Sahlen:2018cku,Kreisch:2018var,Kreisch:2021xzq,Schuster:2019hyl,Zhang:2019wtu,Bayer:2021iyb,Verza:2022qsh,Vielzeuf:2023fqw}, and dark matter properties~\cite{Leclercq:2014pga,Yang:2014upa,Reed:2014cta,Baldi:2016oce,Arcari:2022zul}. This versatility arises from the variety of void-related observables, such as the Alcock-Paczynski test~\cite{Alcock:1979mp,Sutter:2012qwe,Lavaux:2011yh}, redshift-space distortions~\cite{Paz:2013sza,Pisani:2013yxa,Hamaus:2014afa,Hamaus:2015yza,Correa:2021wqw}, weak lensing~\cite{Amendola:1998xu,Melchior:2013gxd,Clampitt:2014gpa,Chuang:2016wqb,Chantavat:2017ysr,DES:2019zaf,Davies:2020udw,Euclid:2022hdx}, baryon acoustic oscillations~\cite{Khoraminezhad:2021bdl}, and the integrated Sachs--Wolfe  effect~\cite{Granett:2008ju,Nadathur:2012ksa,Flender:2012wu,Ilic:2013cn,Cai:2013ik,Kovacs:2015bda,Kovacs:2017hxj,DES:2018nlb,Planck:2015fcm,Dong:2020fqt,Hang:2021kfx,Kovacs:2021mnf}, that are highly sensitive to both the geometry and growth history of the universe. 

Theoretical models of void evolution have historically relied on the spherical collapse model, originally developed for overdense regions (halos)~\cite{Gunn:1972sv,Peebles:1980yev}, and only later adapted to describe underdensities (voids)~\cite{Sheth:2003py,Demchenko:2016uzr,Massara:2018dqb}.  This approach relies on the Newtonian framework for spherical collapse which describes the non-linear evolution of spherically symmetric overdensities (or underdensities) under gravity, using Newton's laws and assuming a pressureless, self-gravitating fluid embedded in an expanding universe. Within the Einstein--de Sitter (EdS) cosmology, this model offers a rare case in which fully analytical solutions can be derived for the  evolution of density perturbations. Importantly, EdS provides a clear analytical criterion for void shell-crossing---the moment when different fluid elements within the void trajectory intersect, marking the onset of strongly non-linear dynamics and defining the formation threshold of a void within the excursion set framework~\cite{Sheth:2003py}.  However, generalising this condition beyond EdS typically requires numerical integration, and remains less well-defined in more realistic cosmological models, such as $\Lambda$CDM or time-dependent DE scenarios.

In this work, we revisit void evolution by employing a hydrodynamical approach, which has previously been applied to the spherical collapse of overdensities~\cite{Pace:2010sn,Abramo:2007iu,Bellini:2012qn,Frusciante:2020zfs}, but has not yet been explored in the context of voids.  The hydrodynamical approach models the evolution of cosmic structures by treating matter as a pressureless fluid governed by the continuity, Euler, and Poisson equations in an expanding universe, providing a physically transparent and extendable framework valid on sub-horizon scales.
To the best of our knowledge, this is the first application of this formalism to model cosmic voids. Our framework has multiple advantages: (i) it provides a physically clear and intuitive description of the evolution of underdense regions, (ii) it is naturally extendable to MG theories, as it is derived directly from the action of the underlying theory, and (iii) most importantly, it allows us to construct a cosmology-dependent mapping between the linear (Lagrangian) and non-linear (Eulerian) evolution of the matter density contrast, a key ingredient for accurate predictions of void statistics~\cite{Sheth:2003py,Verza:2024rbm}.

A second major contribution of this work is the generalisation of the shell-crossing condition beyond EdS. We present a new criterion that enables the numerical computation of the shell-crossing epoch in arbitrary cosmological backgrounds. This result is fundamental for the original formulation of the excursion set formalism for cosmic voids~\cite{Sheth:2003py}, where the shell-crossing threshold is adopted to identify voids in Lagrangian space. While more sophisticated approaches have been developed beyond the original formulation~\cite{Verza:2024rbm}, the shell-crossing condition remains a key ingredient, as it marks the breakdown of the bijective mapping between Lagrangian and Eulerian space. For this reason, accurately determining the shell-crossing threshold remains crucial even in modern techniques used to construct void statistics.

Finally, we apply our formalism to study the evolution of isolated, spherically symmetric voids within the $\Lambda$CDM model, and considering DE cosmologies such as $w_0$CDM and $w_0w_a$CDM. The latter are motivated  by recent DESI observations suggesting potential deviations from a cosmological constant~\cite{DESI:2024uvr,DESI:2024lzq}. We show how our results are relevant in the context of precision cosmology. In the near future, we plan to make publicly available the numerical code associated with this work.

This paper is organized as follows. In section~\ref{Sec:from_Lagrangian_to_Eulerian_space}, we provide an overview of halos and voids in Lagrangian space and their mapping to Eulerian space.  Readers already familiar with this topic may choose to skip this section.  In section~\ref{Sec:Theoretical_approaches_to_spherical_void_evolution},  we study the evolution of isolated spherically symmetric voids: we review the Newtonian framework in section~\ref{Sec:the_spherical_model} and illustrate the application of the hydrodynamical approach to voids in section~\ref{Sec:hydrodynamical_approach}. In section~\ref{Sec:Shell_Crossing}, we derive and present the new shell-crossing criteria suitable for different cosmologies. In section~\ref{Sec:void_evolution}, we present our results on void evolution in different DE cosmologies. Specifically, in section~\ref{Sec:impact_of_cosmology_on_single_void_evolution}, we analyze the impact of the cosmological background on the voids evolution; in section~\ref{Sec:the_linear_to_non_linear_mapping}, we construct the mapping between the linear and non-linear matter density contrast in a cosmology-dependent framework. In section~\ref{Sec:implementation_of_the_shell_crossing}, we present the results of the new shell-crossing criteria across different cosmological scenarios.
Finally, in section~\ref{Sec:Conclusions}, we draw our conclusions.

We close the introduction with table~\ref{tab:definitions} that collects the fundamental definitions that will be employed in the following sections.
\begin{table}[t]
\centering
{
\begin{tabular}{l p{9cm} l l}
\toprule
Symbol & Definition & Space & Dependence  \\
\midrule
$\Delta_{\rm E}$ & 
Mean non-linear matter density contrast & 
Eulerian & $(R_{\rm E},t)$ \\

$\Delta_\mathrm{L}$ & 
Mean linear matter density contrast & 
Lagrangian & $(R_\mathrm{L},t)$   \\

$\delta_{\rm E}$ & 
Non-linear matter density contrast & 
Eulerian & $(\mathbf{r},t)$  \\

$\delta_\mathrm{L}$ & 
Linear matter density contrast & 
Lagrangian & $(\mathbf{r},t)$  \\

$\delta_{\rm v}$ & 
Mapped linear matter density contrast from $\delta_{\rm E}$ at redshift $z$ & 
Lagrangian & $(z,\delta_{\rm E})$  \\

$\delta_{\rm E,sc}$ & 
Non-linear matter density contrast at shell-crossing for a top-hat configuration& Eulerian & (z) \\

$\delta_{\rm v,sc}$ & 
Mapped linear matter density contrast at shell-crossing for a top-hat configuration& 
Lagrangian & $(z)$  \\
\bottomrule
\end{tabular}
\caption{Table of the notation adopted in this paper.}
\label{tab:definitions}
}
\end{table}

\section{From Lagrangian to Eulerian space}
\label{Sec:from_Lagrangian_to_Eulerian_space}
One of the novel contributions of this work concerns the cosmology-dependent mapping between \textit{Lagrangian space}, i.e.~the initial matter density field, $\delta_i$, linearly evolved to the epoch of interest via the growth factor $D$, $\delta({\bf x},z) = \delta_i({\bf x}) D(z)$ and \textit{Eulerian space}, which corresponds to the full non-linear evolved density field.
This mapping can be directly applied to improve the theoretical model of the Void Size Function (VSF)~\cite{Sheth:2003py}. The VSF  describes the number density of voids as a function of their size, typically measured by their effective radius, similarly to the Halo Mass Function (HMF) which models the number density of dark matter halos as a function of their mass~\cite{Bond:1990iw,Press:1973iz}.

Because the VSF quantifies the number density of cosmic voids as a function of their \textit{Eulerian radii}, it requires the mapping from Lagrangian to Eulerian space~\cite{Sheth:2003py,Jennings:2013nsa}, which connects the theoretical prediction, based on initial density fluctuations in Lagrangian space~\cite{Verza:2024rbm}, to the observable quantity, void radii in Eulerian space. In fact, theoretical models of void abundance are formulated in Lagrangian space, where voids are defined as underdense regions in the primordial matter density field~\cite{Sheth:2003py}. These initial conditions (ICs) provide the seeds for the formation of voids and are essential to predict their statistical properties. Observational void catalogues and numerical simulations characterize voids by their Eulerian sizes, their physical radii at a given cosmic time~\cite{Mao:2016faj,Pisani:2019cvo,Neyrinck:2007gy,Nadathur:2013bba,Sutter:2012rts}. To connect theoretical predictions for voids with these observations, it is necessary to evolve the Lagrangian underdensities through cosmic time using a dynamical model of structure formation, typically the spherical model is adopted.
The transition from Lagrangian to Eulerian coordinates encapsulates the non-linear growth of voids. This dynamics is cosmology-dependent, modifying the volume and abundance of voids. In addition, the possible shell-crossing event has to be considered. This event occurs when the trajectory of different fluid elements cross each other, breaking the bijective map from Lagrangian to Eulerian space.

It is worth stressing that, in theoretical and numerical studies of structure formation, Lagrangian perturbation theory (LPT, see, e.g.,~\cite{Bernardeau:2001qr}) is often employed and could provide an interesting complementary perspective to the approach presented here, in particular to go beyond the spherical evolution. However, this framework is not adopted in the present work. 

In the following, we provide an overview of halos and voids in Lagrangian space and their mapping in Eulerian space.

\subsection{Voids and haloes in Lagrangian space}
Let us define the smoothed linearly evolved initial density field at a comoving Lagrangian coordinate $\mathbf{q}$ on the scale $R$, $\delta(\mathbf{q},R)$, as~\cite{Bardeen:1985tr}
\begin{align}\label{eq:densityL}
    \delta(\mathbf{q},R)\,\equiv\,\int\mathrm{d}^3x\,W(\lvert\mathbf{x}\rvert,R)\,\delta(\mathbf{q} + \mathbf{x})\,,
\end{align}
where $\delta (\mathbf{q} + \mathbf{x})$ is the local density field and $W(\lvert\mathbf{x}\rvert,R)$ is a spatial filter that averages the density field over a region of characteristic size $R$, effectively suppressing fluctuations on scales smaller than $R$. Note that we have omitted the time dependence in the above expression. In the excursion-set and peak theory frameworks~\cite{Bardeen:1985tr,Bond:1990iw,Peacock:1990zz,Sheth:1999su,Sheth:2001dp,Paranjape:2011wa}, the identification of halos and voids in Lagrangian space is obtained by applying a threshold criterion to the smoothed initial density field defined in eq.~\eqref{eq:densityL}.\footnote{This picture can be extended by considering other operators of the density field; however, it has been shown that these effects can be approximated with a scale dependent barrier in density only~\cite{Ohta:2004mx,Lazeyras:2015giz,Chiueh:2000yp,Sheth:2001dp,Sheth:2012fc}.}

\paragraph{Halo.} In the standard excursion-set and peak theory frameworks, a dark matter halo characterised by a Lagrangian radius $R_{\mathrm{L}}$ (hereafter the subscript ``L'' is used for quantities defined in the Lagrangian space), is identified at the Lagrangian position $\mathbf{q}$ if $R_{\mathrm{L}}$ corresponds to the largest smoothing scale at which the smoothed density field $\delta(\mathbf{q}, R)$ exceeds the critical linear collapse threshold $\delta_{\rm c}$~\cite{Press:1973iz,Bardeen:1985tr,Bond:1990iw,Peacock:1990zz}.\footnote{Ref.~\cite{Musso:2019zmr} has suggested that the initial energy field, rather than the matter overdensity, may provide a more fundamental variable for halo formation, particularly when the initial profile is not a top-hat. This alternative perspective might also prove interesting in the context of voids.} This criterion is fundamental within the excursion set framework~\cite{Bond:1990iw}, as it provides a systematic method to model the statistics of collapsed structures by comparing the initial density fluctuations to the collapse threshold across multiple spatial scales. This formalism effectively connects the initial density field in Lagrangian space to the non-linear structures observed in Eulerian space, which are at the basis for modeling the HMF and related clustering statistics. 

\paragraph{Void.} Similarly, a void characterised by a Lagrangian radius $R_{\mathrm{L}}$ is identified at the Lagrangian position $\mathbf{q}$ if $R_{\mathrm{L}}$ represents the largest smoothing scale at which the smoothed density field falls below the void formation threshold, that is $\delta_{{\rm v}}$, provided that the collapsing threshold $\delta_{\mathrm{c}}$ is not crossed on any larger smoothing scale. This criterion rigorously accounts for the void-in-void and void-in-cloud processes~\cite{Sheth:2003py} (for a rigorous discussion see~\cite{Paranjape:2012xxx}). Specifically, the requirement that the density crosses $\delta_{{\rm v}}$ at the largest possible scale captures the hierarchical merging of smaller voids into larger voids (void-in-void), whereas the non-crossing of $\delta_{\mathrm{c}}$ at larger scales excludes voids within regions that will collapse (void-in-cloud).

An important conceptual distinction must be made. Halos are gravitationally bound systems that have reached virial equilibrium. As such, a \textit{single} threshold is usually adopted to characterize the collapse: the formation threshold $\delta_\mathrm{c}$ is defined as the value of the linear density contrast corresponding to the formation of a \textit{fully} collapsed structure in Eulerian space.\footnote{Although following the full collapse dynamics is complex, some studies have addressed it using a moving barrier approach~\cite{Paranjape:2011wa,Elia:2011ds,Sheth:2012fc,Ludlow:2011jx,Borzyszkowski:2014xua,Sheth:1999su,Robertson:2008jr,Sheth:2001dp}.} 
The case of voids is different. In the single-stream regime, characteristic of cosmic voids where the matter distribution evolves without shell-crossing and each point in space is associated with a unique velocity field, the threshold $\delta_{\rm v}$ can take \textit{any} negative values, since there always exists a mapping from linear to non-linear space~\cite{Verza:2024rbm}. During void evolution, however, shell-crossing may occur, leading to the breakdown of the single-stream regime. This process marks the stage in void evolution at which distinct matter streams intersect, leading to a transition from coherent, single-stream expansion to a multi-stream flow. This entails the break of the bijective Lagrangian-to-Eulerian map.

When using the spherical model with a top-hat initial profile, as done in computing the VSF, shell-crossing occurs only once, at the boundary between the void and its surrounding environment (see section~\ref{Sec:Shell_Crossing}). This moment marks the breakdown of the spherical model and the onset of a highly non-linear regime that the model is no longer able to describe. Consequently, we claim that the use of thresholds below the shell-crossing point is theoretically inconsistent within the framework adopted for the VSF.

We recall that, in the original proposal of the theoretical model for the VSF~\cite{Sheth:2003py}, the authors adopted the linearly extrapolated value at shell-crossing ($\delta_{\rm v}$) to identify voids, drawing a parallel with the collapse threshold used for halos. However, it is common practice to use an opposite approach, which consists in fixing a non-linear threshold value and using the corresponding linear one, which is always less negative than the extrapolated value at shell-crossing~\citep{Verza:2019tvg,Contarini:2020fdu,Contarini:2022nvd}. 
This is for a twofold reason: on the one hand, the shell-crossing value corresponds to very deep voids, for which the statics is usually poor; on the other hand, the Lagrangian-to-Eulerian map breaks down for values below the shell-crossing threshold, making the theoretical VSF model no more representative of Eulerian voids. Moreover, for thresholds less negative than the shell-crossing value, the mapping becomes steeper and therefore more stable in a data analysis perspective~\cite{Massara:2018dqb}.

For these reasons, we argue that a more robust theoretical treatment of the shell-crossing threshold is necessary.
So far, the linearly extrapolated value at the moment of shell-crossing has only been computed analytically in an EdS universe~\cite{Sheth:2003py}.
In this work, we overcome this limitation by introducing, for the first time, a method to compute the shell-crossing epoch in a fully cosmology-dependent way. The theoretical framework is presented in section~\ref{Sec:Shell_Crossing}, and the results of the numerical implementation are discussed in section~\ref{Sec:implementation_of_the_shell_crossing}.

\subsection{The map from Lagrangian to Eulerian space}
\label{Sec:the_map_from_Lagrangian_to_Eulerian_space}
Here, we do not discuss how the VSF or HMF are constructed in Lagrangian space following the identification of voids or halos, as this lies beyond the scope of the present work. Instead, our focus is on the subsequent step: mapping the VSF or HMF from Lagrangian to Eulerian space, where halos and voids are observed in data or $N$-body simulations. This mapping problem was first examined for voids in~\cite{Sheth:2003py}. For completeness, we briefly review the key ideas in the simplest case, assuming conservation of the number of objects.

\paragraph{Halo.} 
In this case, the number of haloes and their mass are conserved when going from Lagrangian to Eulerian space. In practice, we assume that every object identified as a halo of mass $M_\mathrm{L}(R_\mathrm{L})$ in Lagrangian space corresponds to a halo of the same mass $M_{\rm E}(R_{\rm E})$ in Eulerian space, where the sub-script ``E'' denotes Eulerian quantities. Thus, one can move from one space to the other without making any changes in the HMF.
\paragraph{Void.} 
The case of voids is different, as their statistics are expressed as a function of their radius. Using mass conservation, the Eulerian radius, $R_{\rm E}$, can be derived from the Lagrangian radius, $R_\mathrm{L}$, as
\begin{align}
    R_{\rm E} = \left(1 + \Delta_{\rm E}\right)^{-\frac{1}{3}} R_\mathrm{L} \,,
    \label{Eq:mapping_linear_non_linear_radius}
\end{align}
where $\Delta_{\rm E}$ is defined by spherically averaging the corresponding non-linear density contrast $\delta_{\rm E}$
within a sphere of radius $R_{\rm E}$, as follows
\begin{align}
    \Delta_{\rm E}(R_{\rm E},t) = \frac{3}{4\pi  R_{\rm E}^{3}}\int_0^{R_{\rm E}}{\rm d}ss^{2}\int\mathrm{d}\Omega\,\delta_{\rm E}(\mathbf{s},t)\,,
    \label{Eq:mean_density_contrast}
\end{align}
where $s = \lvert \mathbf{s} \rvert$ is the radial position and $\Omega$ is the solid angle.
The mapping connects a void of Eulerian radius $R_{\rm E}$ and mean density contrast $\Delta_{\rm E}(R_{\rm E})$ to a corresponding Lagrangian region of radius $R_{\rm L}$ and mean linear density contrast $\Delta_{\rm L}(R_{\rm L})$, with the two radii related by mass conservation (eq.~\eqref{Eq:mapping_linear_non_linear_radius}). The mean linear density contrast is defined as
\begin{align}
    \Delta_{\rm L}(R_{\rm L},t) = \frac{3}{4\pi  R_{\rm L}^{3}}\int_0^{R_{\rm L}}{\rm d}s\,s^{2}\int\mathrm{d}\Omega\,\delta_{\rm L}(\mathbf{s},t)\,,
\end{align}
that is, the spherical average of the linear density contrast $\delta_{\rm L}$ within a sphere of radius $R_{\rm L}$. The mapping thus enters the formalism in two distinct ways. First, given a threshold in Lagrangian space $\delta_{\rm v}$, which is needed to compute the VSF, the mapping allows one to determine the corresponding (mean) non-linear density contrast in Eulerian space, $\Delta_{\rm E}$ characterizing observable voids in simulations and real data. Second, it is involved in the construction of the VSF itself, through the transformation of radii from Lagrangian to Eulerian space.

To date, in the literature it has predominantly adopted the simplest approach for this mapping procedure: the so-called \textit{spherical map}, which models void evolution as the expansion of an isolated, spherically symmetric underdensity with an inverse top-hat profile in a cosmological background. The framework is typically applied under the assumption of an EdS universe, where the mapping is analytical.

In section~\ref{Sec:the_linear_to_non_linear_mapping}, we present the first cosmology-dependent implementation of this mapping. Specifically, we introduce a function that, for a void characterized by a non-linear density contrast $\delta_{\rm E}$ at redshift $z$, returns the corresponding linear density contrast under linear evolution, denoted as $\delta_{\rm v}(z, \delta_{\rm E})\,$.

\section{Theoretical frameworks to spherical void evolution}
\label{Sec:Theoretical_approaches_to_spherical_void_evolution}
We study the evolution of isolated, spherically symmetric voids with an initial inverse top-hat density profile in a homogeneous and isotropic universe. This simplified configuration, known as the spherical model, has been widely used to model non-linear structure formation in both over- and underdensities~\cite{Gunn:1972sv,Lilje:1991oiu,Peebles:1980yev,Sheth:2003py,Demchenko:2016uzr}. Although realistic voids are neither spherical nor isolated, this model captures key dynamical features.

In section~\ref{Sec:the_spherical_model}, we review the standard Newtonian approach~\cite{Sheth:2003py}, hereafter dubbed the $R$-based approach, in which the evolution of spherically symmetric underdense regions is described through the physical radius $R(t)$ of the void. Section~\ref{Sec:hydrodynamical_approach} introduces a hydrodynamic formalism, previously applied to collapse~\cite{Pace:2010sn}, here extended to underdensities. 
In section~\ref{Sec:Shell_Crossing}, we propose a new criterion to determine the shell-crossing epoch across general cosmologies, an extension beyond the EdS case. 
This new approach is implemented in a numerical solver, which we plan to make publicly available soon. 

Throughout this section, we refer to the Friedmann-Lema\^{i}tre-Robertson-Walker (FLRW) universe as a flat, homogeneous, and isotropic universe, defined by the line element:
\begin{align}
    \mathrm{d}s^{2} \,=\, g_{\mu\nu}\mathrm{d}x^{\mu}\mathrm{d}x^{\nu} \,=\, -\mathrm{d}t^2 + a^{2}(t)\delta_{ij}\mathrm{d}x^{i}\mathrm{d}x^{j}\,,
\end{align}
where $g_{\mu\nu}$ is the metric tensor, $\mathbf{x}$ is the comoving spatial coordinate vector, $t$ is the cosmic time, $a(t)$ is the scale factor of the universe, and $\delta_{ij}$ is the three-dimensional Kronecker symbol.
\subsection{The \texorpdfstring{$R$}{R}-based approach}
\label{Sec:the_spherical_model}

We assume the energy budget of the universe to be described by pressureless matter (m) and dark energy (DE) only, i.e.,
\begin{align}
    \rho_\text{tot}(t,\mathbf{r}) \,=\, \rho_\mathrm{m}(t,\mathbf{r}) + \rho_\mathrm{DE}(t)\,,
\end{align}
where $\mathbf{r}$ denotes the physical radial coordinate, $\rho_\mathrm{m}$ is the total matter density, including both cold dark matter (CDM) and baryons, and $\rho_\mathrm{DE}$ is the DE density. We note that the matter component has both a background and a perturbation contribution, while DE perturbations are neglected since we assume that they are rapidly smoothed out on scales smaller than the cosmological horizon and do not contribute significantly to structure formation; therefore, DE is treated as a homogeneous component affecting only the background expansion. We also neglect the contribution of radiation, as it is negligible at the redshifts relevant for void analysis. 

Additionally, we consider a spherically symmetric system. From mass conservation, it follows that the mass enclosed within each shell of (Eulerian) radius $R$ remains conserved during expansion, that is, 
\begin{align}
    M(R) = \frac{4 \pi }{3}R^3\,\bar{\rho}_{\mathrm{m}}(t)\left[\,1\,+\,\Delta_{\rm E}(R,t)\,\right] \,=\, \text{const}\,,
    \label{Eq:mass_conservation}
\end{align}
where $M$ is the mass enclosed in the radius $R$ and $\bar{\rho}_{\mathrm{m}}(t)$ is the background matter density. From now on, we denote background quantities with an overbar.\footnote{Let us note that the variable $R$ is written without the subscript E, since in this context we are not referring to a generic Eulerian radius but to the time-evolving radius of the expanding void shells in Eulerian space, as will be explained later in the text.} 

To describe the dynamics of the system, we track the evolution of each individual shell using the variable $R$, representing its physical (Eulerian) radius. Because of spherical symmetry, we can write
\begin{align}
    R \,=\, R\,(t,r_\mathrm{in})\,,
    \label{Eq:variable_in_R}
\end{align}
where $r_{\rm in} = a(t_\mathrm{in})\,r$ denotes the initial physical radius of the shell. Throughout the rest of the paper, we adopt the subscript ``${\rm in}$'' to refer to quantities evaluated at the initial time, which corresponds to the starting point of the system’s dynamics. In practice, $r_{\rm in}$ is a label for each shell. The variable $R$ characterizes the evolution of each shell, capturing deviations from the background expansion induced by the local matter density perturbation. 

Let us now discuss the explicit dependence in eq.~\eqref{Eq:variable_in_R}, and clarify why only $t$ and $r_{\rm in}$ appear. 
This choice of variables is essential for the derivation of the shell-crossing condition in section~\ref{Sec:Shell_Crossing}.
Let us start by examining all variables that could, in principle, enter in eq.~\eqref{Eq:variable_in_R}, beyond $t$ and $r_{\rm in}$. First, we note that any dependence on the initial time $t_\mathrm{in}$ is purely parametric and does not directly affect the dynamics. One might also expect a dependence of $R$ on $\Delta_{\rm E}$; however, eq.~\eqref{Eq:mass_conservation} links $R$ and $\Delta_{\rm E}$, making them interdependent and interchangeable in describing the system's evolution. Finally, while $R$ can also depend on the mean initial density contrast $\Delta_{\rm E}(r_{\rm in},t_{\rm in})$, this quantity itself depends on $t_\mathrm{in}$ and $r_{\rm in}$. Hence, expressing $R$ as a function of $t$ and $r_{\rm in}$ is appropriate.

Adopting a pseudo-Newtonian cosmology approach~\cite{Abramo:2007iu,Abramo:2008ip,Creminelli:2009mu,Pace:2022fam,Pace:2018xqy,DelPopolo:2012jm}, the evolution of each shell with radius $R(t, r_{\rm in})$ can be written as  
\begin{align}
    \frac{\ddot{R}}{R} \,=\, -\frac{4\pi G}{3}\sum_{j}\,\left[\,\rho_{j}(R,t) + 3\,p_{j}(R,t)\,\right]\,,
    \label{Eq:Newton_equation}
\end{align}
where $G$ is the Newtonian gravitational constant, overdots denote derivatives with respect to cosmic time, i.e.~$\dot{R} \equiv \partial R / \partial t$ and $j$ is a label for matter (m) and dark energy (DE), respectively.  
 
In this paper, we make two further assumptions to solve the dynamics explicitly. First, we take the initial matter density profile to be an inverse top-hat profile. The profile reads 
\begin{align}
    \delta_{\rm E}(t_\mathrm{in}, r_{\rm in}) \,=\, 
    \begin{cases}
        \delta_\mathrm{v,in} & \text{for } r_{\rm in} \leq r_{\rm v,in} \\
        0 & \text{for } r_{\rm in} > r_{\rm v,in}
    \end{cases}\,,
    \qquad\quad
    \Delta_{\rm E}(r_{\rm in},t_\mathrm{in}) = 
    \begin{cases}
        \delta_{\rm v,in} & \text{for } r_\mathrm{in} \leq r_{\rm v,in} \\
        \delta_\mathrm{v,in}\left(\frac{r_{\rm v,in}}{r_{\rm in}}\right)^3 & \text{for } r_{\rm in} \geq r_{\rm v,in}
    \end{cases}\,,
    \label{Eq:initial_top_hat_delta}
\end{align}
where $r_{\rm v,in}$ is the initial (in) radius of the void (v) and $\delta_{\rm v,in}$ is the initial value of the matter density contrast inside the void. While the approach is easily extended to any DE model, here for concreteness we adopt the Chevallier-Polarski-Linder (CPL) parametrization~\cite{Chevallier:2000qy,Linder:2002et} for the equation of state (EoS) of DE, $w_{\rm DE}(a)\equiv \bar{p}_\mathrm{DE}(t)/\bar{\rho}_\mathrm{DE}(t)$, 
\begin{align}
    w_{\rm DE}(a) \,=\, w_0 + w_a\left(1-a\right)\,,
\end{align}
where $w_0$ and $w_a$ are constant. The former is the present-day value of the DE equation of state, the latter is the derivative of $w_{\rm DE}$ with respect to the scale factor
at present time.

In the case of a top-hat, all shells within the initial void radius ($r_{\rm in} < r_{\rm v,in}$) evolve identically, as the density is constant within each sphere (see, e.g.,~\cite{Goldberg:2003bw,Wagner:2014aka,Dai:2015jaa}). Their evolution follows that of a FLRW universe with $\rho_\mathrm{m} = \bar{\rho}_\mathrm{m}(t)\left[1+\delta_\mathrm{m}(t)\right]$. Hence, we can focus directly on the outermost shell, $r_{\rm in} = r_{\rm v,in}$. 

Eq.~\eqref{Eq:Newton_equation} admits an analytical solution only in the EdS case (see ref.~\cite{Sheth:2003py} and appendix~\ref{Sec:Appendix_EdS}). For other cosmologies, and even for the $\Lambda$CDM model, it cannot be solved analytically and must instead be integrated numerically (see, e.g.,~\cite{Demchenko:2016uzr}). This equation remains valid until shell-crossing, beyond which the model breaks down, as it is no longer possible to track the evolution of individual shells. We defer the theoretical description of shell-crossing to section~\ref{Sec:Shell_Crossing}.

\subsection{Hydrodynamical approach}
\label{Sec:hydrodynamical_approach}
We propose an alternative framework for modeling the evolution of isolated spherically symmetric cosmic voids embedded in a homogeneous and isotropic cosmological background, based on Newtonian hydrodynamics. This approach employs the matter density contrast as the primary dynamical variable. Although this formalism is well established in the context of spherical collapse for halos formation~\cite{Padmanabhan:1996qwe, Abramo:2007iu, Pace:2010sn, Pace:2017qxv}, it has not been systematically applied to underdense regions, despite the fact that the underlying dynamical equations are formally identical.

We adopt the Newtonian gauge for the perturbed spatially flat FLRW metric in Cartesian coordinates, i.e.
\begin{align}
    \mathrm{d}s^2 \,=\, -\left[1+2\Psi(\mathbf{x},t)\right]\mathrm{d}t^2 + a^2(t)\left[1-2\Phi(\mathbf{x},t)\right]\delta_{ij}\mathrm{d}x^i\mathrm{d}x^j\,,
\end{align}
where $\Phi(\mathbf{x},t)$ and $\Psi(\mathbf{x},t)$ are the two gravitational potentials. We work under the same assumptions of the previous section, i.e.~spherical symmetry, matter and DE are the only components, and DE is treated as a background component. We assume a perfect fluid form for the stress-energy tensor of pressureless matter ($p_\mathrm{m}=0$), i.e.
\begin{align}
    T_{\mu\nu}\,=\,\rho_\mathrm{m}\,u_\mu u_\nu\,,
    \label{Eq:stress_energy_tensor}
\end{align}
where $\rho_\mathrm{m}$ and $u_\mu$ are the energy density and the four-velocity of the fluid, respectively. By perturbing the Einstein equations in the Newtonian limit, we obtain from the 00 and $ij$ components~\cite{Pace:2010sn} 
\begin{align}
         \nabla^2_{\mathbf{x}}\Psi   & = 4\pi G a^2 \bar{\rho}_{\mathrm{m}}\delta_{\mathrm{E}}\,, \label{Eq:00_perturbed_equations} \\
         \nabla^2_\mathbf{x}(\Psi+\Phi)   & = 8\pi G a^2 \bar{\rho}_{\mathrm{m}} \delta_{\mathrm{E}}\,,\label{Eq:ij_perturbed_equations}
\end{align}
where $\nabla_\mathbf{x}$ stands for derivatives with respect to the coordinate $\mathbf{x}$. From eqs.~\eqref{Eq:00_perturbed_equations} and \eqref{Eq:ij_perturbed_equations}, we have $\Phi=\Psi$, which is the result of the absence of anisotropic stresses in eq.~\eqref{Eq:stress_energy_tensor}~\cite{Kodama:1984ziu}.

By perturbing the conservation law of the stress-energy tensor, i.e.~$\nabla_\mu T^{\mu\nu}=0$, in the Newtonian limit, we get~\cite{Pace:2010sn} 
\begin{align}
    \dot{\delta}_{\rm E} + \left(1+\delta_{\rm E}\right)\nabla_{\mathbf{x}}\cdot\vec{u} &\,\,=\,\,0\,,
    \label{Eq:energy_conservation}\\
    \frac{\partial\vec{u}}{\partial t} + 2H\vec{u} + \left(\vec{u}\cdot\nabla_\mathbf{x}\right)\vec{u} + \frac{1}{a^2}\nabla_\mathbf{x}\Phi &\,\,=\,\,0\,,
    \label{Eq:momentum_conservation}
\end{align}
where $\vec{u}$ is the spatial component of the comoving peculiar velocity and $H(t)$ is the Hubble parameter. Now, by combining the divergence of eq.~\eqref{Eq:momentum_conservation}, the time derivative of eq.~\eqref{Eq:energy_conservation}, and eq.~\eqref{Eq:00_perturbed_equations} under the assumption of spherical symmetry, we get the evolution equation for the non-linear matter density~\cite{Pace:2010sn} in $e-$fold time $x=\ln{a}$
\begin{align}
    \delta_{\rm E}^{''} \,+\, \left(2 + \frac{H^{'}}{H} \right)\,\delta_{\rm E}^{'} \,-\, \frac{4}{3} \,\frac{(\delta_{\rm E}^{'})^2}{1\,+\,\delta_{\rm E}} - \frac{3}{2} \,\Omega_\mathrm{m}\, (1 + \delta_{\rm E})\,\delta_{\rm E}\,=\,0\,,
    \label{Eq:non_linear_evolution_equation}
\end{align}
where $'\equiv\partial/\partial x$ and $\Omega_{\rm m}(x)$ is the matter density parameter. The linearized version of eq.~\eqref{Eq:non_linear_evolution_equation} for the linear matter density contrast $\delta_\mathrm{L}$ is given by
\begin{align}
    \delta_\mathrm{L}^{''} \,+\, \left(2 \,+\, \frac{H^{'}}{H} \right)\,\delta_\mathrm{L}^{'} \,-\, \frac{3}{2} \,\Omega_\mathrm{m}\,\delta_\mathrm{L} \,=\, 0\,.
    \label{Eq:linear_evolution_equation}
\end{align}
Eqs.~\eqref{Eq:non_linear_evolution_equation} and \eqref{Eq:linear_evolution_equation} are valid for any spherically symmetric initial matter density profile.\footnote{We note that the spatial dependence enters only parametrically through the initial profile.} 
Relating the evolution equations for the local density contrasts, $\delta_{\rm E}$ and $\delta_{\rm L}$, to those for the mean quantities, $\Delta_{\rm E}$ and $\Delta_{\rm L}$, is in general a non-trivial task, even within spherical symmetry. For the linear equation, this correspondence can be established explicitly: integrating eq.~\eqref{Eq:linear_evolution_equation} over the radius yields the evolution equation of the mean contrast $\Delta_{\rm L}$. In the non-linear case, however, quadratic terms such as $\delta_{\rm E}^2$ prevent a straightforward connection between the local and the integrated evolution. Establishing such a general mapping is beyond the scope of this work. In our analysis (see section~\ref{Sec:void_evolution}), we restrict to the inverse top-hat configuration, where the situation simplifies considerably, as the local and mean quantities inside the void coincide, as we now discuss.

When considering an inverse top-hat, as in eq.~\eqref{Eq:initial_top_hat_delta}, all points inside the void ($r_{\rm in}<r_{\rm v,in}$) share the same initial conditions. 
Having the same initial conditions, therefore, implies that the solution at any time $t$ is the same for all points within the void. 
In this case, local and mean quantities coincide, allowing one to use either description interchangeably. 
At any cosmic time $t$ we thus have
\begin{align}
    \delta_{\rm E}(r,t) \, = \, \Delta_{\rm E}(r,t)\,, \quad \text{ for all }\, r < R(t,r_{\rm in,v})\,,
\end{align}
where $R(t,r_{\rm in,v})$ denotes the time evolution of the outermost shell of the void, as determined by the formalism introduced in section~\ref{Sec:the_spherical_model}. 
In the hydrodynamical formalism, one could in principle describe the void dynamics by following a single point inside the underdense region, since the evolution equations are local. However, in spherical symmetry with an initial inverse top-hat configuration, that single point can be naturally associated with a shell of radius $R$, whose mean density is $\Delta_{\rm E}$. In this way, we can provide a direct connection between the hydrodynamical description and the $R$-based formulation. 
In the inverse top-hat case, all inner shells evolve identically, so it is sufficient to follow a single one, typically the outermost shell $R_{\rm v}\equiv R(t,r_{\rm in,v})$.

Using mass conservation, it can then be shown that solving eq.~\eqref{Eq:non_linear_evolution_equation} for the density contrast $\delta_{\rm E}$ (and subsequently obtaining $\Delta_{\rm E}(t,R_{\rm v})$) is equivalent to solving eq.~\eqref{Eq:Newton_equation} for $R_{\rm v}$. For non–top-hat profiles, however, this equivalence is not straightforward to establish, as it requires an explicit mapping between local and mean quantities.

In the following, we \textit{always} consider the evolution of spherical voids with an inverse top-hat profile for all our results. 
Although it is customary in the literature to describe the dynamics of the voids in terms of the mean density contrast $\Delta_{\rm E}$, we instead adopt the notation $\delta_{\rm E}$, consistent with the hydrodynamical formalism employed throughout this work. 
For a top-hat configuration, the two quantities coincide, and the results presented in the following sections are therefore unaffected by this choice of notation.

The hydrodynamical approach offers several advantages over the $R$-based formulation: 
\begin{itemize}
    \item It offers a more transparent physical interpretation of void dynamics by directly tracing the evolution of the matter density within the void.
    \item From a theoretical standpoint, the equations of motion (EoMs) can be derived directly from the action of the theory, which, in the framework adopted in this work, is given by
    \begin{align}
        S \,=\, \int \mathrm{d}^4x\,\sqrt{-g} \left( \frac{1}{16\pi G} R \,+\, \mathcal{L}_\mathrm{m} \,+\, \mathcal{L}_\mathrm{DE} \right)\,,
    \end{align}
    where $g$ is the determinant of $g_{\mu\nu}$, $R$ is the Ricci scalar, and $\mathcal{L}_\mathrm{m}$ and  $\mathcal{L}_{\rm DE}$ denote the Lagrangian densities for matter and DE, respectively. While both the hydrodynamical and $R$-based approaches are formally equivalent, the latter relies on pseudo-Newtonian reasoning to obtain eq.~\eqref{Eq:Newton_equation}. In contrast, the hydrodynamical framework derives all relevant equations directly from the action, offering a more transparent and theoretically consistent formulation.
    \item 
    This formalism becomes particularly relevant when extending the analysis to MG theories. In most MG scenarios, the primary modification arises in the Poisson equation, which deviates from its standard form given in eq.~\eqref{Eq:00_perturbed_equations}, while the overall derivation structure remains unchanged.

    \item This method provides, for the first time, a transparent and rigorous framework for deriving the cosmology-dependent mapping between Lagrangian and Eulerian space.
    The mapping can be derived using $R$ as the dynamical variable, but it is an involved procedure that requires changing variables from $R$ to $\delta_{\rm E}$. To obtain the linear equation, i.e., eq.~\eqref{Eq:linear_evolution_equation}, one must change variables to derive eq.~\eqref{Eq:non_linear_evolution_equation} from eq.~\eqref{Eq:Newton_equation} and then linearize it. Thus, we believe that the hydrodynamic approach offers a simpler connection to linear theory being a significant advancement toward a more physically motivated and general description of structure formation across cosmological models.
\end{itemize}
In the following, we solve numerically eqs.~\eqref{Eq:non_linear_evolution_equation} and~\eqref{Eq:linear_evolution_equation}, which are second-order differential equations. Therefore, two conditions are necessary. The first one, as standard practice for the spherical collapse model~\cite{Bellini:2012qn}, is set at \textit{early time} by assuming that the integration starts during the matter-dominated era, when matter perturbations are linear, i.e.~$\delta_{\rm E}=\delta_{\rm L}$ and the decaying mode (of matter perturbations) can be disregarded. This is equivalent to set
\begin{align}
    \delta_{\rm E}^{'}(x_{\rm in}) = \delta_{\rm E}(x_{\rm in}) = \delta_{\rm v,in}\,,
\end{align}
where $\delta_{\rm v,in}$ is not fixed a priori. Indeed, it is determined by imposing a \textit{late time} condition on the non-linear matter density contrast. For example, if we are interested in studying a void with $\delta_{\rm E}(z=0) = -0.4$, we perform a shooting procedure on $\delta_{\rm v,in}$ until the solution satisfies this late-time condition. Here, we have outlined the methodology used to set the conditions required to solve the system dynamics, without discussing the details of the procedure and the underlying assumptions, which are instead addressed in appendix~\ref{Sec:the_numerical_integration}. In this work, we take as the starting point for the integration of the equations $a_{\rm in} = 10^{-7}$.

\subsection{The shell-crossing condition}
\label{Sec:Shell_Crossing}
In this section, the issue of shell-crossing in the spherical model is examined through a brief qualitative overview, followed by a review of the standard analytical derivation in the EdS cosmology.

As a void expands, matter flows outward from the center, leading to a  density gradient in which the inner shells become increasingly underdense compared to the outer ones. This gradient causes the inner shells to move outward more rapidly than the outer shells. Eventually, the inner shells overtake the outer ones, resulting in shell overlap and the breakdown of the single-stream regime, a phenomenon known as shell-crossing. A similar process occurs in the case of spherical collapse. After the turnaround point, the outer shells fall inward more rapidly than the inner ones, and shell-crossing arises when they overtake the inner shells during infall toward the center.

\begin{figure}[t]
    \centering
    \includegraphics[width=0.7\textwidth]{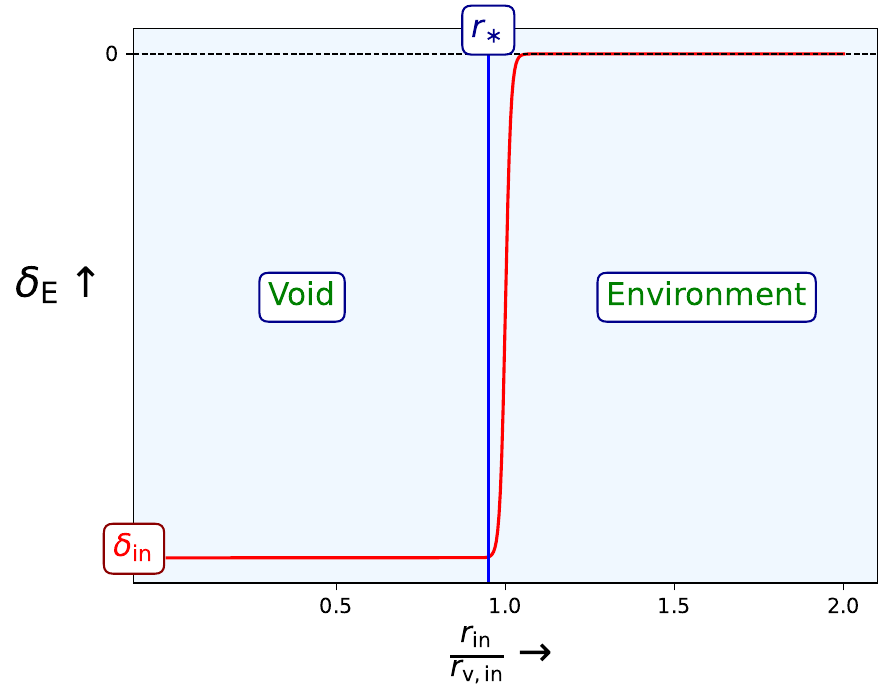}
    \caption{The continuous density profile $\delta_{\rm E}$ is shown as a function of the initial physical radial coordinate $r_{\rm in}$ for a spherical void with initial radius $r_{\rm v,in}$. The void is embedded in a cosmological background environment, defined as the region where matter perturbations are absent.
    Here, $r_\ast$ marks the radius at which the matter density contrast $\delta_{\rm E}$ ceases to be constant and starts to increase.}
    \label{Fig:step_function}
\end{figure}

Now, we specialize to the case in which the initial matter density profile of the void is an inverse top-hat profile and we provide a qualitative description of the shell-crossing. 
Due to the idealized characteristics of this profile, an initial consideration of a smooth density distribution, as depicted in figure~\ref{Fig:step_function}, is warranted to accurately characterize the dynamics. The analysis can then be extended by examining the limiting behavior as the profile approaches discontinuity. In the plot, we show an isolated spherical void with initial radius $r_{\rm v,in}$ and $\delta_\mathrm{v,in} < 0$ in a ``cosmological background environment''. In this context, we refer to the latter as the region where the matter and DE densities are at their background values and perturbations are absent.
In figure~\ref{Fig:step_function}, $r_\ast$ marks the radius beyond which the density stops being constant and begins to increase. Up to $r_\ast$, the dynamics follows that of a Friedmann universe with lower matter density~\cite{Goldberg:2003bw,Wagner:2014aka,Dai:2015jaa}, and no shell-crossing occurs.
However, the shells $r_\ast$ and $r_\ast\,+\,\mathrm{d}r_{\rm in}$ cross almost immediately as the void begins to evolve. From that point on, we expect successive crossings to take place between adjacent shells, until the outermost shell (of the void) overlaps with the background environment (see, e.g.,~\cite{Massara:2018dqb}). This marks the \textit{epoch of shell-crossing}. Throughout the rest of the paper, we define the shell-crossing time as the moment when the outermost void shell crosses with the background environment. Using an inverse top-hat profile leads to a loss of resolution of the crossing between the outermost shells of the void at the void boundary because of the unphysical discontinuity in the matter density. Nevertheless, as will be demonstrated, it is still possible to compute the moment when the outermost shells of the void cross with those of the environment, which is the only moment of interest.

The next step involves translating the conceptual discussion into a rigorous mathematical framework. To begin with, we consider two shells at the initial time $t_\mathrm{in}$ such that
\begin{align}
    R\big(r_{\mathrm{in_1}},\, t_\mathrm{in}\big) > R\big(r_{\mathrm{in_2}},\, t_\mathrm{in}\big)
    \quad
    \text{with} 
    \quad 
    r_{\mathrm{in_1}} > r_{\mathrm{in_2}}\,,
\end{align}
where $r_\mathrm{in_1}$ and $r_\mathrm{in_2}$ are labels assigned to each shell in order to track their evolution (see section~\ref{Sec:the_spherical_model}).  
We say that a shell-crossing has occurred if 
\begin{align}
    R\big(r_{\mathrm{in_1}}, t \big) 
    < 
    R\big(r_{\mathrm{in_2}}, t\big)\,
    \quad
    \text{with}
    \quad
    t > t_\mathrm{in}\,.
\end{align}
We are now interested in computing the time at which shell-crossing occurs for two shells that are infinitesimally close at $t_\mathrm{in}$, i.e.~
\begin{align}
    R_1 \,= \,R\left(\bar{r}, t_\mathrm{in}\right)\,,\quad  R_2 \,=\, R\left(\bar{r}+\mathrm{d}\bar{r}, t_\mathrm{in}\right)\,.
    \label{Eq:infinitesimally_close_shells_initial_time}
\end{align}
Shell-crossing occurs when the separation between the two shells vanishes at a fixed instant in time. This is equivalent to request~\cite{Blumenthal:1992ert}
\begin{align}
    \begin{cases}
        \mathrm{d}R & = \,0\,,\\
        \mathrm{d}t & = \, 0\,.
    \end{cases}\,\,
    \label{Eq:shell_crossing_condition_analytical}
\end{align}
This is the condition usually adopted in the literature to compute the epoch of shell-crossing in the case of the EdS model (see, e.g.,~\cite{Blumenthal:1992ert,Sheth:2003py}). This condition can only be applied if a parametric analytical solution exists. We review this computation in appendix~\ref{Sec:Appendix_EdS}, where we also show how to compute the epoch of shell-crossing in the case of a top-hat profile, and demonstrate how the issue of the discontinuity can be bypassed to compute what we previously defined as the epoch of shell-crossing. 
The condition in eq.~\eqref{Eq:shell_crossing_condition_analytical} cannot be used when solving the EoMs numerically, as required in other cosmological scenarios. 

We present a comprehensive formalism to compute the \textit{shell-crossing epoch numerically}, enabling precise determination of the time at which matter streams intersect during void or halo evolution. This approach facilitates a detailed analysis of the non-linear dynamics governing structure formation and provides a robust framework applicable to a wide range of cosmological models.
To this end, the system’s dynamics must be expressed in terms of the variable $R$ in eq.~\eqref{Eq:variable_in_R}, as the shell-crossing condition applies directly to the shells. Then, we first derive the shell-crossing criterion within the 
$R$-based framework and subsequently demonstrate its implementation in the hydrodynamical formalism.

Let us consider two infinitesimally close shells at the initial time $t_\mathrm{in}$, as in eq.~\eqref{Eq:infinitesimally_close_shells_initial_time}. Using the explicit dependence of the variable $R$ given in eq.~\eqref{Eq:variable_in_R}, we can translate the requirement that ``the separation between the two shells vanishes at a fixed instant'' into
\begin{align}
    0 \,&=\, \left.\mathrm{d}R \right|_{t} \,=\, \left.\left[\frac{\partial R(t,r_{\rm in})}{\partial r_{\rm in}}\,  \mathrm{d}r_{\rm in} + \frac{\partial R(t,r_{\rm in})}{\partial t} \, \mathrm{d}t\right]\right|_{t} \,=\, \frac{\partial R(t,r_{\rm in})}{\partial r_{\rm in}}\,  \mathrm{d}r_{\rm in}  \,,
    \label{Eq:shell_crossing_condition_R_1}
\end{align}
where $\left.[\ldots]\right|_{t}$ denotes the relative quantity evaluated at fixed time. 

When working with an inverse top-hat profile,  eq.~\eqref{Eq:shell_crossing_condition_R_1} must be adapted to account for the discontinuity in the radial derivative of $R$ at $r_{\rm v,in}$.
Taking the left-hand derivative is equivalent to asking when two shells cross in a Friedmann universe, in which case the derivative is identically zero. Therefore, to determine the moment of shell-crossing, we take 
\begin{align}
    \lim_{\varepsilon\,\to\,0^+}\,\frac{R\,(t,r_{\rm v,in}+\varepsilon)-R\,(t,r_{\rm v,in})}{\varepsilon} \,=\,0\,. 
    \label{Eq:Shell_Crossing_R_R_right_hand_derivative}
\end{align}
In section~\ref{Sec:Consistency_checks_with_the_EdS_model}, we implement this condition in our numerical solver and show that, in an EdS universe, eq.~\eqref{Eq:Shell_Crossing_R_R_right_hand_derivative} reproduces the analytical result.  

Although derived within the $R$-based description, the condition can be adapted to the hydrodynamic formalism.  
To handle the discontinuity in the matter density profile at the void boundary, we work with the mean density contrast $\Delta_{\rm E}$ instead of $\delta_{\rm E}$.\footnote{For $r_{\rm in} \leq r_{\rm v,in}$, $\Delta_{\rm E}$, and $\delta_{\rm E}$ are identical, so using one or the other as the dynamical variable makes no difference for the evolution.} Differentiating with respect to $r_{\rm in}$ the mass conservation equation, where the constant is fixed by evaluating the mass at the initial time and imposing eq.~\eqref{Eq:shell_crossing_condition_R_1}, yields
\begin{align}
    R^3\,\bar{\rho}_{\mathrm{m}}\,\frac{\partial }{\partial r_{\rm in}} \Delta_{\rm E}(R,t) \,=\, r_{\rm in}^2\,\bar{\rho}_{\rm m,in} \,\left\{3\,\left[1+\Delta_{\rm E}(r_{\rm in},t)\right]+r_{\rm in}\,\frac{\partial }{\partial r_{\rm in}} \Delta_{\rm E}(r_{\rm in},t_\mathrm{in})\right\}\,.
\end{align}
Since we use $\Delta_{\rm E}(R,t)$ as the dynamical variable, we must rewrite derivatives with respect to $r_{\rm in}$ in terms of $\Delta_{\rm in}(r_{\rm in})\equiv\Delta_{\rm E}(r_{\rm in},t_{\rm in})$. This requires a relation between $r_{\rm in}$ and $\Delta_{\rm in}$, which we obtain by differentiating eq.~\eqref{Eq:mean_density_contrast} with respect to $r$, which reads
\begin{align}
    \frac{\mathrm{d}}{\mathrm{d}r}\Delta_{\rm E}(r,t) \,=\, 3\,\frac{\Delta_{\rm E}(r,t)}{r}\left[\frac{\delta_{\rm E}(r,t)}{\Delta_{\rm E}(r,t)}-1\right]\,.
    \label{Eq:derivative_delta}
\end{align}
Assuming a top-hat profile as the initial density profile, we can evaluate eq.~\eqref{Eq:derivative_delta} at $t_\mathrm{in}$ as
\begin{align}
    \frac{\mathrm{d} \ln \Delta_\mathrm{in}}{\mathrm{d} \ln r_{\rm in}} = 
    \begin{cases}
        \,0 & \text{for } r_{\rm in} < r_{\rm v,in} \\
        \,-3 & \text{for } r_{\rm in} > r_{\rm v,in}
    \end{cases}\,.
    \label{Eq:derivative_Delta_i_r_i}
\end{align}
As in eq.~\eqref{Eq:shell_crossing_condition_analytical}, we take the right-hand derivative in eq.~\eqref{Eq:derivative_Delta_i_r_i}, following the same reasoning.  
Thus, the shell-crossing condition reads
\begin{align}
    \frac{\mathrm{d} \Delta_{\rm E} }{\mathrm{d} \Delta_{\rm in}}  \,=\, -\frac{(1+\Delta_{\rm E})}{(1+\Delta_{\rm in})\,\Delta_{\rm in}} \,.
    \label{Eq:Shell_crossing_Delta}
\end{align}
Despite its simplicity, this equation enables a precise determination of the shell-crossing time in any cosmological scenario, though the calculation must be carried out numerically. This is one of the main results of this work.

\section{Void evolution}
\label{Sec:void_evolution}
We apply the hydrodynamical formalism to study the void dynamics in the standard cosmological model, $\Lambda$CDM, and a $w_0w_a$CDM cosmological model   characterized by a CPL parametrization of the DE equation of state. While the $R$-based method has previously been used for the $\Lambda$CDM~\cite{Demchenko:2016uzr}, this work presents the first implementation of the hydrodynamical framework in this context. 
In appendix~\ref{Sec:Consistency_checks_with_the_EdS_model}, we present a series of consistency checks performed to validate the numerical implementation of both the $R$-based and hydrodynamical approaches against the analytically solvable EdS case, providing a robust validation of the numerical solver.

In this section, we present three key results that we summarize in the following: 
\begin{itemize}
    \item \textit{Impact of cosmology on single void evolution.} In section~\ref{Sec:impact_of_cosmology_on_single_void_evolution}, we present the evolution of an isolated void in a uniform cosmological background. We assess the impact of a background DE component—either a cosmological constant or a dynamical form of DE—on the evolution of cosmic voids by varying the equation of state parameters $w_0$ and $w_a$. We also explore the role of the present-day matter density parameter, $\Omega_{\mathrm{m},0}$, showing, once again, that cosmic voids could be used as an independent cosmological probe to constrain this parameter (see, e.g.,~\cite{Contarini:2022nvd,Hamaus:2020cbu,eBOSS:2020nuf}). These are the only background parameters that can vary in eqs.~\eqref{Eq:non_linear_evolution_equation} and~\eqref{Eq:linear_evolution_equation}, and they all enter the perturbation dynamics through their effect on the time-dependent matter parameter $\Omega_{\rm m}(t)$.
    \item \textit{The linear to non-linear mapping.}
    In section~\ref{Sec:the_linear_to_non_linear_mapping}, we present, for the first time, the results for the cosmology-dependent mapping $\delta_{\rm v}(z,\delta_{\rm E})$ from Lagrangian to Eulerian space discussed in section~\ref{Sec:from_Lagrangian_to_Eulerian_space}. 
    \item \textit{Shell-crossing.}
    In section~\ref{Sec:implementation_of_the_shell_crossing}, we compute $\delta_{\rm v}(z,\delta_{\rm E,sc})$ by determining $\delta_{\rm E}$ dynamically, requiring that voids reach shell-crossing, i.e.~$\delta_{\rm E} =\delta_{\rm E,sc}$. This is the \textit{first implementation} of such a procedure in the literature. 
The computational approach adopted in this context mirrors the standard method used to compute the collapse threshold $\delta_\mathrm{c}$ in the spherical model (see, e.g.,~\cite{Pace:2010sn}).
\end{itemize}
In our analysis, we explore a range of values for the present-day matter (\text{CDM + baryons}) density parameter $\Omega_{\mathrm{m},0} \in [0.2, 1.0]$, the DE EoS parameter $w_0 \in [-2.0, -0.5]$, and its ``evolution'' parameter $w_a \in [-2.0, 0.1]$. The results are presented in a series of plots, each showing the variation of one parameter at a time, while keeping all others fixed. When varying $w_0$ or $w_a$, we fix $\Omega_{\mathrm{m},0} = 0.32$. On the other hand, when varying $\Omega_{\mathrm{m},0}$, we assume the DE to be a cosmological constant, i.e.~$w_0 = -1$ and $w_a = 0$. In the following, we refer to the EdS model as the case where $\Omega_{\mathrm{m},0} = 1$, corresponding to a matter-dominated universe with no DE component.  
We refer to the $\Lambda$CDM model as the scenario with $\Omega_{\mathrm{m},0} = 0.32$, and  characterized by $w_0 = -1$ and $w_a = 0$.

The ranges for $w_0$ and $w_a$ are motivated by observational constraints from the combination of DESI and CMB data presented in eq.~(25) of~\cite{DESI:2025zgx}. However, not all parameter values considered here fall strictly within the current $68\%$, $95\%$ or $99\%$ confidence regions. This broader exploration is intentional: it allows us to study how cosmic voids respond to a wide class of dynamical DE models.

Should the DESI measurements be confirmed by other cosmological probes, our analysis could help quantify the impact of dynamical DE on void evolution and statistics. More broadly, this framework may also prove useful for high-precision cosmological surveys (as \textit{Euclid}), where void statistics could serve as an independent probe to test and constrain models of dynamical DE.

Throughout this section, we \textit{always} consider spherical voids with an inverse top-hat density profile.
As discussed in section~\ref{Sec:hydrodynamical_approach}, in this case the mean non-linear (linear) density contrast $\Delta_{\rm E}$ ($\Delta_{\rm L}$) and the local density contrast $\delta_{\rm E}$ ($\delta_{\rm L}$) are identical inside the void.
In the literature, the evolution of an isolated spherical void in a homogeneous and isotropic background is usually described in terms of mean quantities ($\Delta_{\rm E}$ and $\Delta_{\rm L}$).
However, since the local and mean quantities coincide in this setup, one can equivalently use either of them.
For consistency with the hydrodynamical formalism introduced (for the first time here for voids) and employed in this work, we use $\delta_{\rm E}$ and $\delta_{\rm L}$ throughout the analysis. The results remain unchanged if expressed in terms of $\Delta_{\rm E}$ and $\Delta_{\rm L}$, as the two quantities are identical within the void.

\subsection{Impact of cosmology on single void evolution}
\label{Sec:impact_of_cosmology_on_single_void_evolution}
In this section, we present results for the evolution of a single, isolated, spherically symmetric void across different cosmological backgrounds. Halos and voids exhibit different sensitivities to dark energy: while halos collapse and enclose progressively smaller regions, thereby reducing the impact of smooth components such as dark energy or massive neutrinos, voids expand and probe increasingly larger volumes, which makes them progressively more affected by such components. This further justifies going beyond the EdS approximation.

\begin{figure}[t!]  
    \centering\includegraphics[width=1\textwidth]{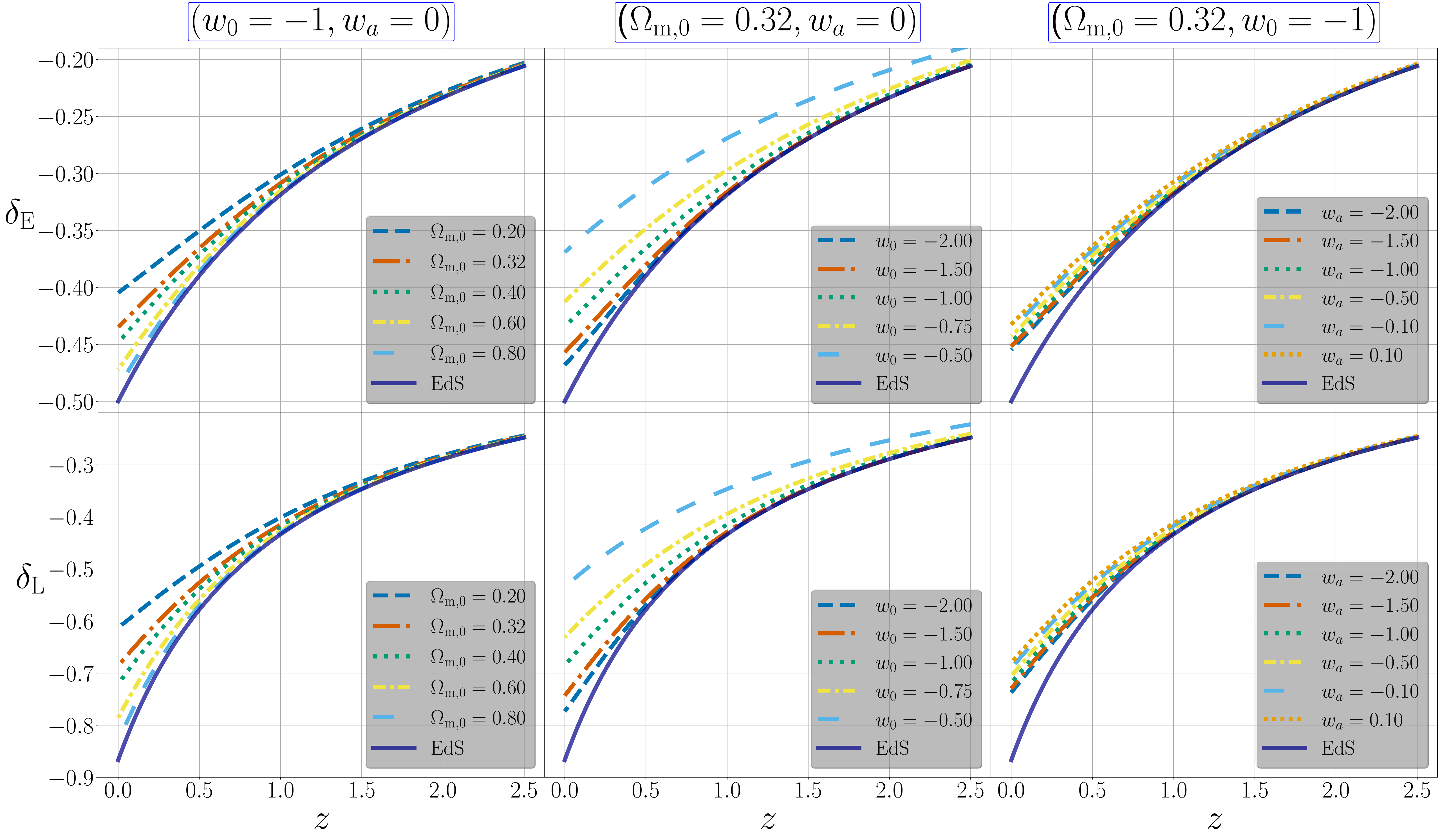}
    \caption{Evolution of $\delta_{\rm E}$ (top panels) and $\delta_{\rm L}$ (lower panels) as functions of redshift in the range $z \in [0, 2.5]$ for different cosmological models. In the left column, we fix $(w_0 = -1, w_a = 0)$ and vary the present-day matter density parameter $\Omega_{\mathrm{m},0}$. In the central columns, we fix $(\Omega_{\rm m,0} = 0.32, w_a = 0)$ while varying $w_0$. In the right column, we fix $(\Omega_{\rm m,0} = 0.32, w_0 = -1)$ while varying $w_a$. The initial conditions ($\delta_{\rm v,in}$ at $x_{\rm in}$) for all the solutions shown are identical and correspond to those that, in an EdS model, lead to $\delta_{\rm E}(z=0) = -0.5$. }\label{Fig:single_void_different_scenarios}
\end{figure}

We analyze the evolution of a single void reaching $\delta_{\rm E} = -0.5$ at redshift $z = 0$ in an EdS universe. While this choice is arbitrary, this value can reflect voids detectable in galaxy surveys. We then extract the corresponding initial conditions from the EdS evolution, i.e.~$
    \delta_{\rm v,in}(x_\mathrm{in})  \,=\, \delta_{\rm v,in}^{\mathrm{EdS}}\,,$
and we use this value to evolve the void across different cosmologies. This setup allows us to isolate the effect of the background expansion history on the non-linear growth of the void.
We present the results in figure~\ref{Fig:single_void_different_scenarios}. 

In the first column, we analyze the effect of varying $\Omega_{\mathrm{m},0}$ and fix the background evolution to be the flat $\Lambda$CDM one.
As expected, the larger the value of $\Omega_{\mathrm{m},0}$, the more rapidly the void evacuates its matter content—i.e., the emptier it becomes. This is because increasing the background matter density enhances the strength of gravitational attraction, thereby accelerating structure formation. This effect can be traced back to the source term in eq.~\eqref{Eq:non_linear_evolution_equation}. The same effect can also be understood from a complementary perspective: the presence of DE slows down the growth of structures by weakening the gravitational pull that drives structure formation. Both interpretations are equivalent, as they reflect the suppression of $\Omega_{\rm m}(x)$ in the source term in eq.~\eqref{Eq:non_linear_evolution_equation}, ultimately leading to a slower growth of underdensities.  Although this is not shown in the plots, we emphasize that all the solutions converge to the EdS one at higher redshifts, as all cosmological background models do. This holds true for each plot in figure~\ref{Fig:single_void_different_scenarios}.

\begin{figure}[t]  
    \centering\includegraphics[width=0.75\textwidth]{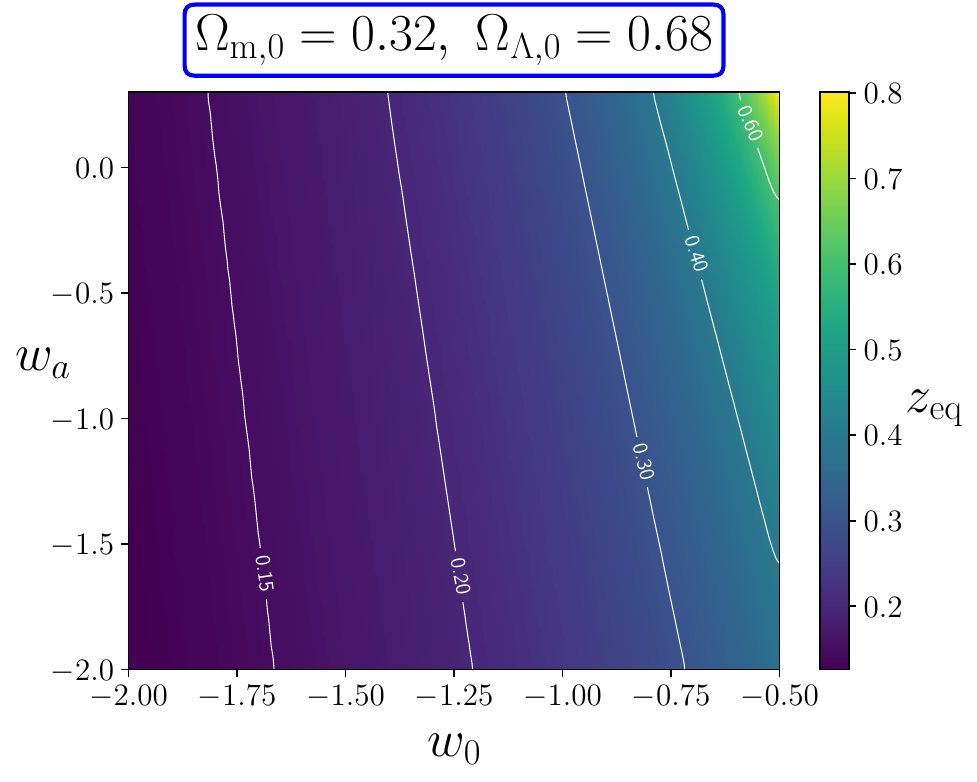}
    \caption{The redshift of matter–dark energy equality, $z_{\rm eq}$, defined by $\rho_{\rm m}(z_{\rm eq}) = \rho_{\rm DE}(z_{\rm eq})$, as a function of $w_0$ and $w_a$, while keeping $\Omega_{\rm m,0} = 0.32$ fixed.}
    \label{Fig:z_eq}
\end{figure}

In the second column, we fix the background parameters to $(\Omega_{\mathrm{m},0} = 0.32, w_a = 0)$ and explore the impact of varying $w_0$ on void evolution. For comparison, we also display the EdS model. The interpretation of the results mirrors that discussed in the previous case, as the effect is again driven by the role of DE in suppressing structure formation. Specifically, varying the value of $w_0$ changes the redshift $z_{\rm eq}$ of the matter–dark energy equality, defined as  $\rho_{\rm m}(z_{\rm eq}) = \rho_{\rm DE}(z_{\rm eq})$. As shown in figure~\ref{Fig:z_eq}, less negative values of $w_0$ (keeping $w_a$ fixed) lead to earlier matter–dark energy equality, corresponding to higher values of $z_{\rm eq}$. In turn, this results in a longer period during which DE dominates the expansion, leading to a stronger suppression of the growth of cosmic structures. This trend is reflected in the void evolution shown in figure~\ref{Fig:single_void_different_scenarios}, where increasingly negative $w_0$ values yield an emptier void.

In the third column, we vary $w_a$ while setting $(\Omega_{\mathrm{m},0}=0.32\,\,w_0=-1)\,.$  As shown in figure~\ref{Fig:z_eq}, increasing the value of $w_a$ (while keeping $w_0$ fixed) shifts the matter–dark energy equality to higher redshifts, resulting in a more significant suppression of structure growth. This is consistent with what is observed in figure~\ref{Fig:single_void_different_scenarios}. However, the differences with respect to the $\Lambda$CDM case, when keeping $w_0 = -1$ fixed and varying $w_a$, remain small and are barely visible in figure~\ref{Fig:single_void_different_scenarios}. Thus, we now investigate whether different choices of $w_0$ amplify the effect of varying $w_a$, potentially leading to more pronounced deviations from $\Lambda$CDM.

\begin{figure}[t]  
    \centering\includegraphics[width=1\textwidth]{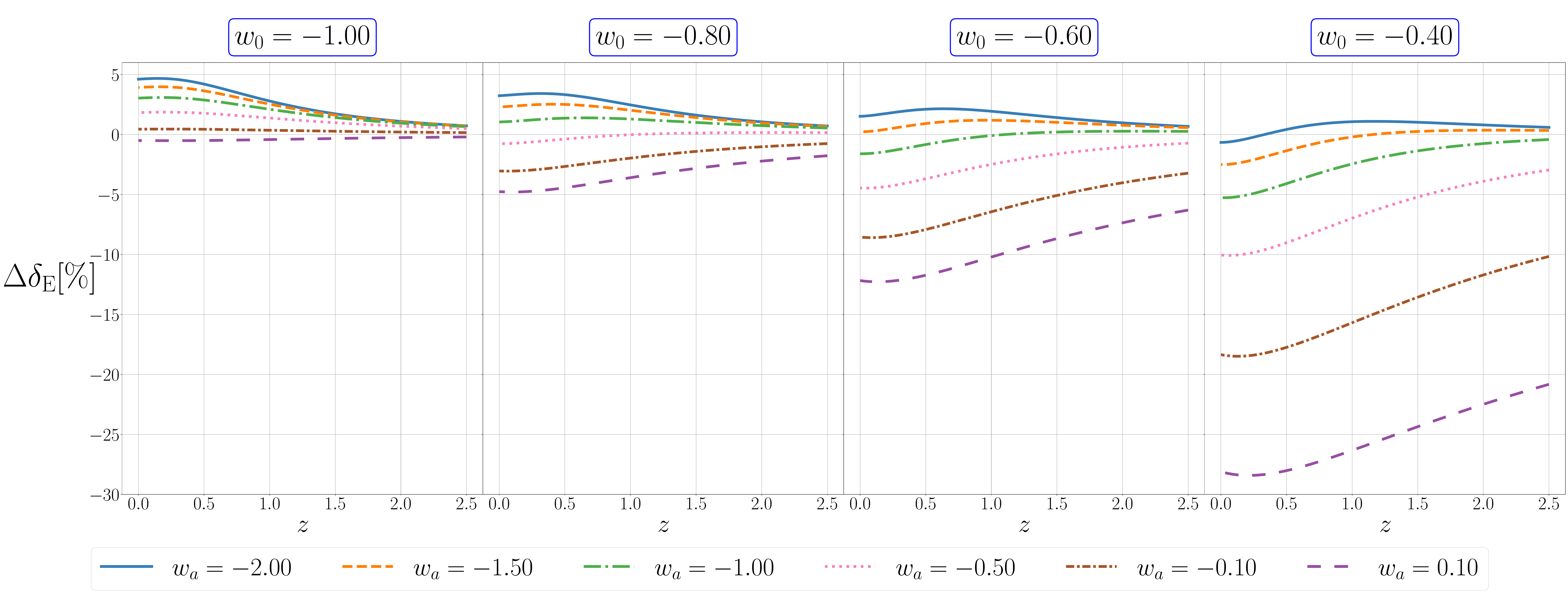}
    \caption{Percentage relative difference in the non-linear matter density contrast $\delta_{\rm E}$ between $w_0w_a$CDM and $\Lambda$CDM models, plotted as a function of redshift. In all panels, the matter density is fixed to $\Omega_{\mathrm{m},0} = 0.32$, while $w_a$ is varied over the range $[-2, 0.1]$. Each column corresponds to a different choice of $w_0$, with values $-1$, $-0.8$, $-0.6$, and $-0.4$ from left to right.}
    \label{Fig:impact_w_a}
\end{figure}

To this end, in figure~\ref{Fig:impact_w_a}, we plot the percentage relative differences between the solutions in $w_0w_a$CDM and those in a $\Lambda$CDM universe, i.e.
\begin{align}
    \Delta\delta_{\rm E}[\%]  = \frac{\delta_{\rm E} - \delta_{\rm E,\Lambda  CDM}}{\delta_{\rm E,\Lambda CDM}}\times 100\,.
\end{align}
In all panels, we fix $\Omega_{\mathrm{m},0} = 0.32$ and explore the impact of varying $w_a$ in the range $[-2, 0.1]$, for four different values of $w_0$: $-1$, $-0.8$, $-0.6$, and $-0.4$ in the first, second, third, and fourth columns, respectively. Our results suggest that the evolution of the total matter density contrast ($\delta_{\rm E}$) inside voids is quite sensitive to the DE EoS, with differences reaching up to 20--30$\%$. The differences are particularly large for less negative values of $w_0$. 

Although not explicitly shown in the figure, all curves tend to zero at high redshift, reflecting the fact that $w_0w_a$CDM and $\Lambda$CDM models become observationally indistinguishable in the matter-dominated regime. Here, not all the curves are monotonic, as they would be if the comparison were performed with respect to EdS. In that case, the deviations would grow monotonically with time; since DE becomes increasingly dominant at late times, the growth of voids would be progressively slower compared to EdS, resulting in an ever-increasing discrepancy. However, in this case, the reference model is $\Lambda$CDM, which itself includes a cosmological constant. As a result, the behavior of the percentage difference depends on the relative timing of $z_{\rm eq}$ in $\Lambda$CDM and in the $w_0w_a$CDM models being considered. The curves reflect the different evolution of the DE component in the two models, which varies with redshift. The final outcome is determined by the interplay between these effects, which we do not explore in further detail here.

To conclude this section, we aim to assess the impact of different cosmological models on the late-time dynamics of void evolution. To this end, we consider an alternative setup in which the ``late-time'' condition, introduced in section~\ref{Sec:hydrodynamical_approach}, is fixed at a sufficiently high redshift, deep in the matter-dominated era, where differences among cosmologies are expected to be negligible. In particular, we set this redshift to $z = 99$, a standard choice for the starting point of $N$-body simulations. We then evolve the non-linear solutions down to redshift zero to quantify how the background cosmology affects the final outcome. The integration of the non-linear evolution equation, i.e.~eq.~\eqref{Eq:non_linear_evolution_equation}, always starts at $z_\mathrm{in} = 10^7$, as specified in section~\ref{Sec:hydrodynamical_approach}. To assess the impact of the initial integration point on the final results, see appendix~\ref{Sec:the_numerical_integration}.

\begin{figure}[t!]      
\centering\includegraphics[width=1\textwidth]{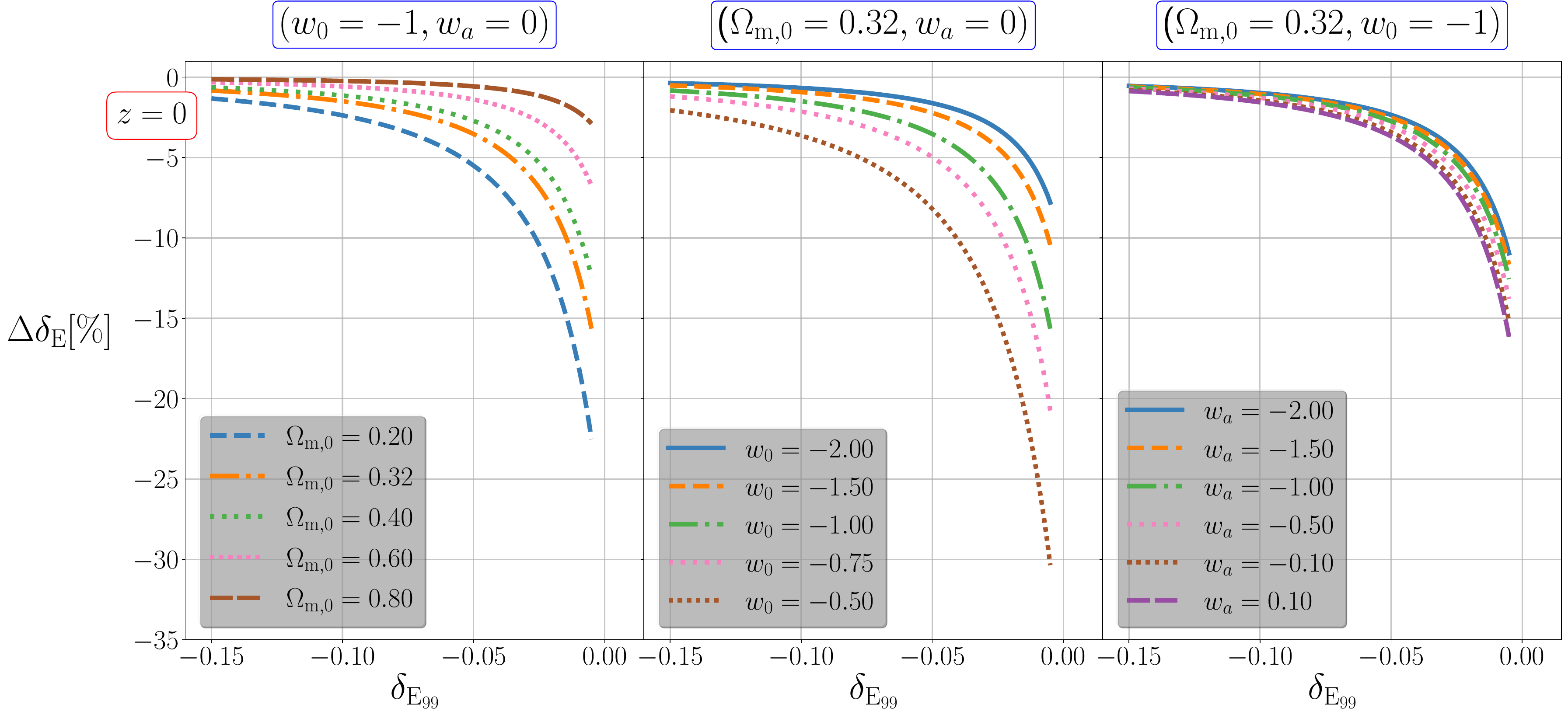}
    \caption{Percentage relative difference in the non-linear density contrast $\delta_{\rm E}$ at $z = 0$ between various cosmological models and the EdS reference. In all cases, the ``late-time'' condition for solving eq.~\eqref{Eq:non_linear_evolution_equation} is set at $z = 99$, where $\delta_{\rm E}(z=99)$ is sampled in the range $[-0.15\,,-0.001]$.  The three columns show the effect of varying $\Omega_{\mathrm{m},0}\in[0.2,0.8]$ (left), $w_0\in[-2,-0.5]$ (center), and $w_a\in[-2,0.1]$ (right), with all other parameters held fixed. }
    \label{Fig:comparison_ics_LCDM}
\end{figure}

More specifically, we fix the non-linear density contrast at $z = 99$ within a given range, e.g.
\begin{align}
    \delta_{\rm E}(z=99) \,\equiv\,\delta_{\rm E_{99}}\in [-0.05\,,-0.001]\,.
\label{Eq:range_ICs}
\end{align}
For each value in this range, we determine the corresponding initial condition $\delta_{\rm v,in}$ at $x_\mathrm{in}$ by means of a shooting procedure, such that the evolved solution matches the desired $\delta_{\rm E}$ at $z = 99$. These initial conditions are then used to evolve the system down to redshift zero, allowing us to quantify the impact of different cosmological models on the late-time evolution. 
\begin{figure}[t!]
    \centering
    \includegraphics[width=0.85\linewidth]{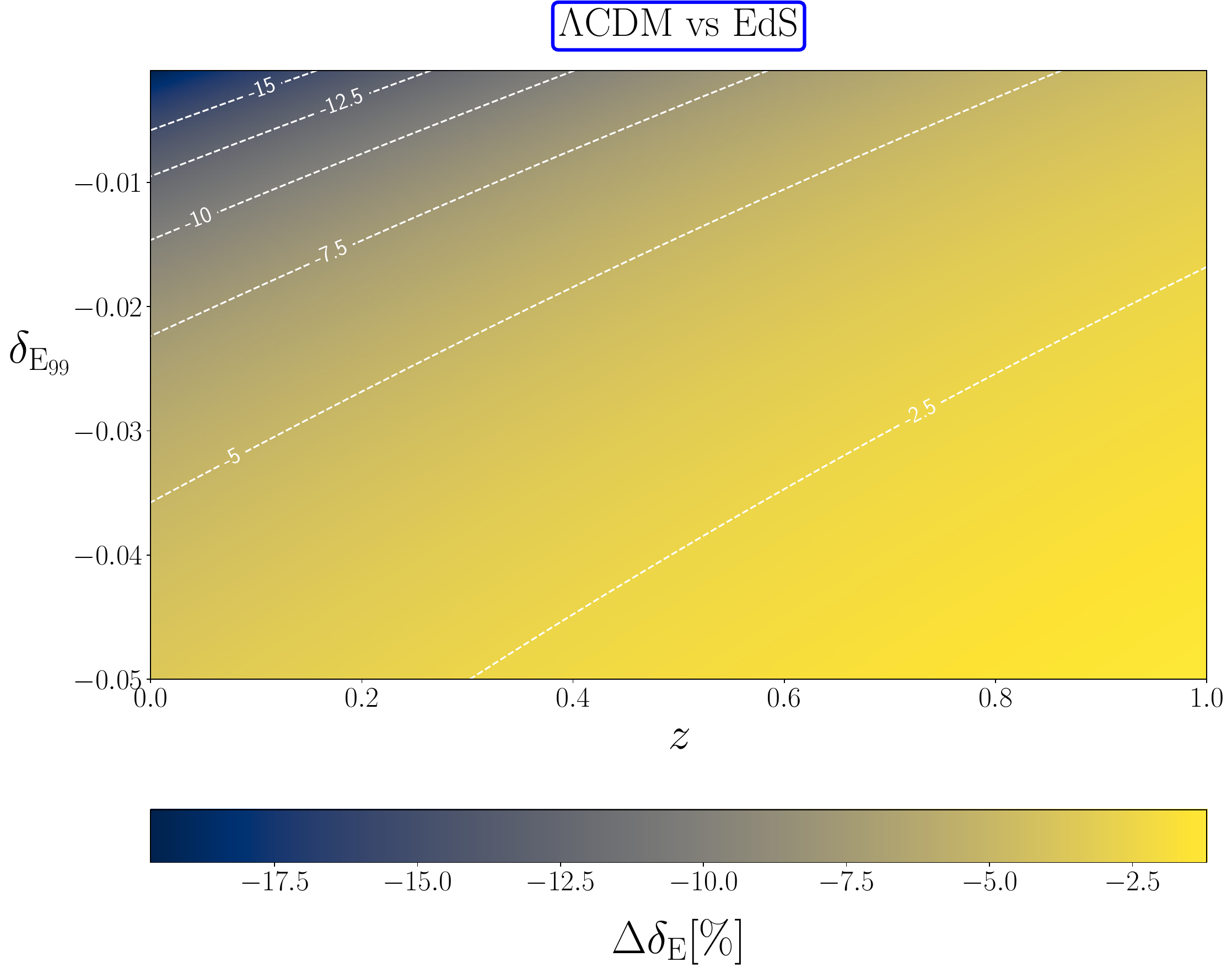}
    \caption{Relative percentage difference between the non-linear density contrast $\delta_{\rm E}(z)$ in $\Lambda$CDM and in an EdS universe, shown as a function of redshift in the interval $z \in [0,1]$. In this setup, we fix the ``late-time'' condition (see section~\ref{Sec:hydrodynamical_approach}) at $z = 99$ by sampling $\delta_{\rm E}(z=99)$ within the range $[-0.05\,,-0.001]$, as specified in eq.~\eqref{Eq:range_ICs}, and then evolve each case down to redshift zero. All differences are computed with respect to the EdS solution, using the same conditions at $z = 99$.}
    \label{Fig:map_ics_lcdm_eds}
\end{figure}
We present the results in Figures~\ref{Fig:map_ics_lcdm_eds} and~\ref{Fig:comparison_ics_LCDM}. 

In figure~\ref{Fig:map_ics_lcdm_eds}, we focus on the $\Lambda$CDM model. We plot the percentage relative difference in the redshift range $z\in[0,1]$ between the non-linear solutions obtained by evolving each condition $\delta_
{\rm E_{99}}$ in eq.~\eqref{Eq:range_ICs} in $\Lambda$CDM and the corresponding solutions in the EdS case. This quantity is defined as
\begin{align}
    \Delta\delta_{\rm E}[\%](\delta_{\rm E_{99}},z) \,=\, \frac{\delta_{\rm E}^{(\text{model})}(\delta_{\rm E_{99}},z) - \delta_{\rm E}^{(\text{EdS})}(\delta_{\rm E_{99}},z)}{\delta_{\rm E}^{(\text{EdS})}(\delta_{\rm E_{99}},z)}\times 100\,,
    \label{Eq:percentage_diff_map_ics}
\end{align}
where $\delta_{\rm E}^{(\text{model})}$ and $\delta_{\rm E}^{(\text{EdS})}$ denote the non-linear density contrasts at redshift $z$ in the given cosmological model and in the EdS case, respectively. 
As expected, the differences between $\Lambda$CDM and EdS become smaller at higher redshifts, reflecting the fact that the $\Lambda$CDM background converges to the EdS limit in the deep matter-dominated regime. Starting from the same conditions at $z=99$, the departure (between the $\Lambda$CDM and EdS solutions for $\delta_{\rm E}$) grows progressively towards lower redshifts, where the cosmological constant starts to affect the dynamics more significantly.
We also find that the percentage differences are always negative, indicating that the evolved non-linear density contrast in $\Lambda$CDM is always less negative than in EdS, i.e.~
\begin{align}
    \delta_{\rm E}^{\Lambda{\rm CDM}}(\delta_{\rm E_{99}},z) \,>\, \delta_{\rm E}^{\rm EdS}(\delta_{\rm E_{99}},z)\,.
\end{align}
This is consistent with the expected slower structure formation in $\Lambda$CDM.
Deviations up to $20\%$ are observed for the shallowest voids (those with $\delta_{\rm E_{99}}\simeq -0.001$), indicating a greater sensitivity to the background expansion with respect to the deepest ones (those with 
$\delta_{\rm E_{99}}\simeq -0.05$). In contrast, the evolution of deeper voids is less dependent on the choice of cosmological model. This can be interpreted as evidence that the impact of the cosmological background on void evolution is much stronger in the linear or quasi-linear regime, whereas the dynamics of deeply non-linear voids are predominantly governed by their internal gravitational potential. In particular, the shallower voids in our setup—i.e., those characterized by smaller underdensities at $z = 99$—are still in the linear or mildly non-linear regime when dark energy becomes dynamically relevant, and are therefore more sensitive to the background expansion history. In contrast, deeper voids undergo most of their growth during the matter-dominated era, reaching a non-linear stage before dark energy takes over, which renders their subsequent evolution largely insensitive to the global cosmological model.

In figure~\ref{Fig:comparison_ics_LCDM}, we assess the impact of varying $\Omega_{\rm m,0}$, $w_0$, and $w_a$ on the late-time dynamics of void evolution. We plot the percentage relative difference defined in eq.~\eqref{Eq:percentage_diff_map_ics} at $z = 0$ for different cosmological backgrounds. The plots follow the same structure of figure~\ref{Fig:single_void_different_scenarios}: in the first column, we vary $\Omega_{\mathrm{m},0}\in[0.2,0.8]$; in the second, $w_0\in[-2,-0,5]$; and in the third, $w_a\in[-2,0.1]$. In all cases, the ``late-time'' condition is fixed at $z = 99$ with $\delta_{\rm E}(z=99)$ sampled in the range $[-0.15\,,-0.001]$. 
The fact that all differences are negative confirms that structure growth is systematically slower in $w_0w_a$CDM than in EdS. Moreover, the larger percentage differences introduced by shallower voids reflect their stronger dependence on the background expansion history, consistent with the interpretation discussed above.
The impact of the cosmological parameters varied in these plots—$\Omega_{\mathrm{m},0}$, $w_0$, and $w_a$—can, also in this case, be interpreted in terms of the redshift of matter–dark energy equality, $z_{\rm eq}$. Decreasing $w_0$ or $w_a$, or increasing $\Omega_{\rm m,0}$, shifts $z_{\rm eq}$ to lower redshifts, thereby shortening the time during which DE influences the evolution. As a result, the solutions tend to converge toward the EdS case, and the percentage differences are correspondingly reduced.
Notably, deviations as large as $30\%$ are observed in the explored parameter space, confirming that the evolution of the non-linear matter density contrast $\delta_{\rm E}$ inside voids is highly sensitive to the DE equation of state. Interestingly, the largest deviations occur for parameter values compatible with current DESI constraints. If these measurements are confirmed, cosmic voids could provide another valuable tool to assess the impact of the DE sector on the LSS of the universe.

\subsection{The linear to non-linear mapping}
\label{Sec:the_linear_to_non_linear_mapping}
In this section, we present the results regarding the impact on the map from Lagrangian (linear) to Eulerian (non-linear) space when changing the cosmological background. Before proceeding, we recall the definition of the mapping (see section~\ref{Sec:the_map_from_Lagrangian_to_Eulerian_space}).
The mapping is defined as a function that, given a void with a certain non-linear matter density contrast $\delta_{\rm E}$ at a specific redshift $z$, returns the corresponding linearly extrapolated value $\delta_{\rm v}(z, \delta_{\rm E})$ — that is, the value the void would have if it had evolved according to linear theory. This is the \textit{physical} interpretation of the map.

\textit{Mathematically}, $z$ and $\delta_{\rm E}$ can be arbitrarily specified, provided that $\delta_{\rm E}$ remains above the shell-crossing threshold at that redshift (see section~\ref{Sec:from_Lagrangian_to_Eulerian_space}). Once these two values are set, they are interpreted as describing a void with density contrast $\delta_{\rm E}$ at redshift $z$, and the goal is to determine what its density contrast would be at the same redshift if it had followed linear evolution (eq.~\eqref{Eq:linear_evolution_equation}).
Varying $z$ while keeping $\delta_{\rm E}$ fixed corresponds to considering voids with the same non-linear density contrast at different cosmic times. Once again, as discussed in section~\ref{Sec:the_map_from_Lagrangian_to_Eulerian_space}, values of $\delta_{\rm E}$ more negative than the shell-crossing threshold are theoretically inconsistent and are therefore excluded from the analysis. In all the following plots, we account for this by imposing a minimum threshold $\delta_{\rm E} \geq -0.75$, which we have verified to be above the shell-crossing value $\delta_{\rm E,sc}$ at all redshifts considered. 

Numerically, the mapping is constructed as follows. We fix a redshift, e.g., $z = 1$, and a target value for the non-linear matter density contrast (at that redshift), e.g., $\delta_{\rm E} = -0.4$.
We then perform a shooting method to determine the initial condition $\delta_{\rm v,in}$ at $x_{\rm in}$ such that the non-linear evolution (see eq.~\eqref{Eq:non_linear_evolution_equation}) of the void leads to the desired final density contrast:
\begin{align}
    \delta_{\rm E}(x_{\rm in}) \,=\, \delta_{\rm v,in} \quad\longrightarrow\quad \delta_{\rm E}\,(z = 1) \,=\, -0.4\,.
\end{align}
Once this initial condition is determined, we evolve it forward using the linear evolution equation (see eq.~\eqref{Eq:linear_evolution_equation}) up to the same redshift $z = 1$.
This procedure yields the linearly extrapolated density contrast associated with the same final non-linear configuration.

\begin{figure}[t!]  
    \centering\includegraphics[,width=1\textwidth]{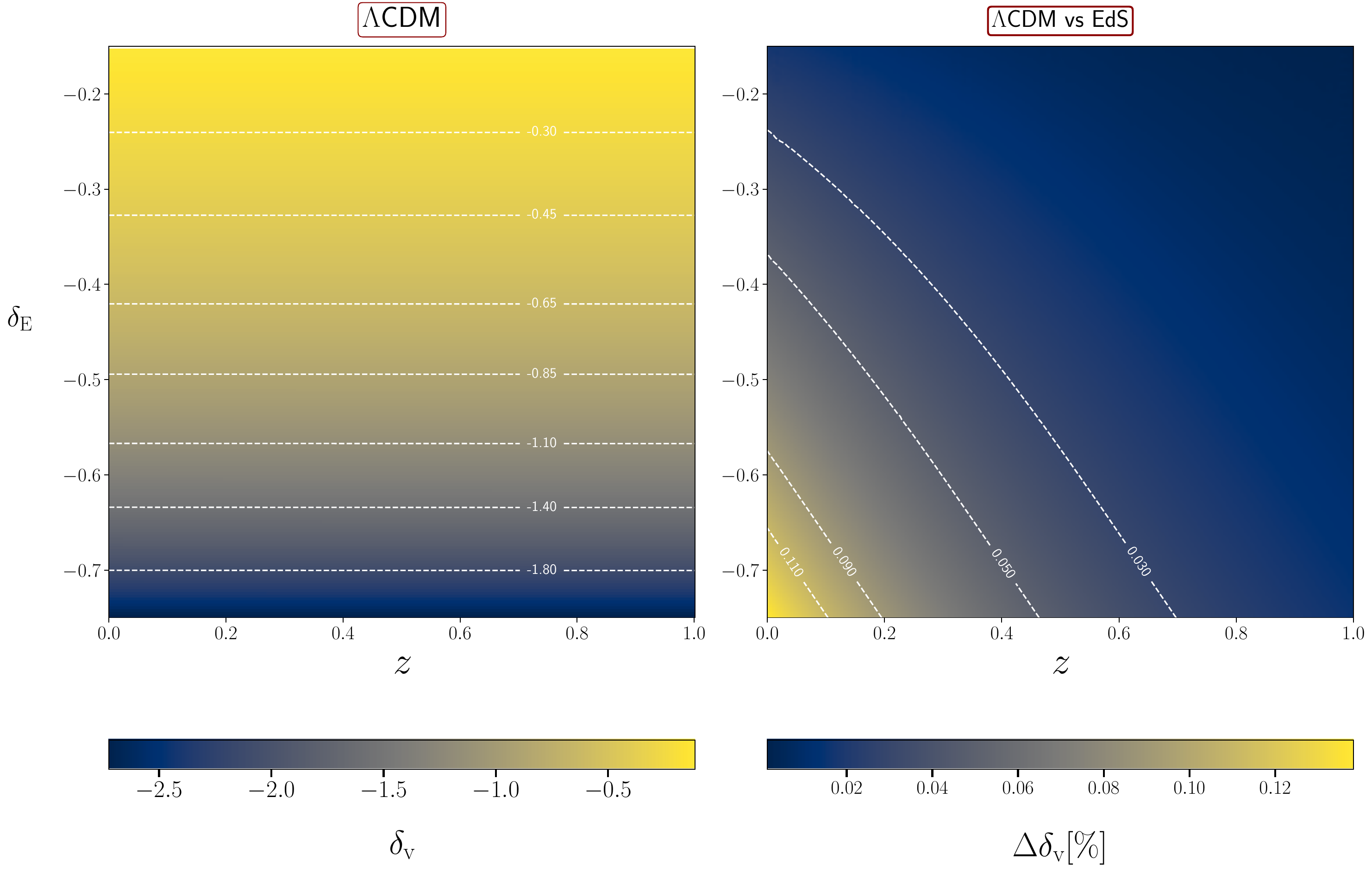}
    \caption{Left panel: we present the non-linear to linear mapping, i.e.~$\delta_{\rm v}(z,\delta_{\rm E})$ for the $\Lambda$CDM model over the range $z\in[0,1]$ and $\delta_{\rm E}\in[-0.75,\,-0.15].$ Right panel: we present the relative percentage difference, i.e.~$\Delta\delta_{\rm v}[\%]$, between the $\Lambda$CDM model and the EdS one, using the same range for $z$ and $\delta_{\rm E}$.}
    \label{Fig:map_LCDM_vs_EdS}
\end{figure}
To date, the EdS mapping is commonly used in the literature (see, e.g.,~\cite{Verza:2024rbm,Jennings:2013nsa,Euclid:2022qtk}). Here, we present results for cosmologies with a CPL parametrization of the DE equation of state. We begin by showing, in figure~\ref{Fig:map_LCDM_vs_EdS}, a color map of the mapping function in the $\Lambda$CDM model. In the left panel, we present the mapping over the range $z \in [0, 1]$ and $\delta_{\rm E} \in [-0.75,\,-0.15]$. In the right panel, we display the relative percentage difference between the $\Lambda$CDM model and the EdS one, i.e.,
\begin{align}
    \Delta\delta_{\rm v}[\%](z,\delta_{\rm E})\,\equiv\,\frac{\delta_{\rm v}^{(\Lambda\mathrm{CDM})}(z,\delta_{\rm E}) - \delta_{\rm v}^{(\mathrm{EdS})}(z,\delta_{\rm E})}{\delta_{\rm v}^{(\mathrm{EdS})}(z,\delta_{\rm E})}\times100\,.
    \label{Eq:percentage_difference}
\end{align}
As shown in figure~\ref{Fig:map_LCDM_vs_EdS}, the differences between the two mappings are small, typically below the percent level.
Despite these small deviations, one can still expect an impact on the theoretical predictions for the VSF. This situation is analogous to the case of spherical collapse, where sub-percent-level differences in the collapsed threshold $\delta_{\rm c}$ between EdS and $\Lambda$CDM translate into percent-level variations in HMF (see, e.g.,~\cite{Pace:2010sn}). It is therefore reasonable to expect a similar sensitivity in void statistics.
Importantly, any systematic effect on the VSF induced by the mapping could, in principle, be exploited for cosmological inference. In this perspective, the mapping itself could become a potential tool to discriminate between different cosmological models, particularly in the context of high-precision void cosmology. A more detailed exploration of the redshift and $\delta_{\rm E}$ dependence of the mapping will be presented in the following.

Then, we examine how the mapping $\delta_{\rm v}(z, \delta_{\rm E})$ depends on the underlying cosmological model. Specifically, we investigate its sensitivity to variations in the parameters $\Omega_{\mathrm{m},0}$, $w_0$, and $w_a$ within the $w_0w_a$CDM framework. This analysis requires evaluating the mapping at either fixed redshift or fixed non-linear density contrast. Thus, in addition to isolating the effect of each parameter, this approach also allows us to revisit and better visualize the small but non-negligible differences observed earlier between $\Lambda$CDM and EdS, while providing a theoretical discussion of the origin of the $z$ and $\delta_{\rm E}$ dependence in the mapping.
\begin{figure}[t]  
    \centering\includegraphics[width=1\textwidth]{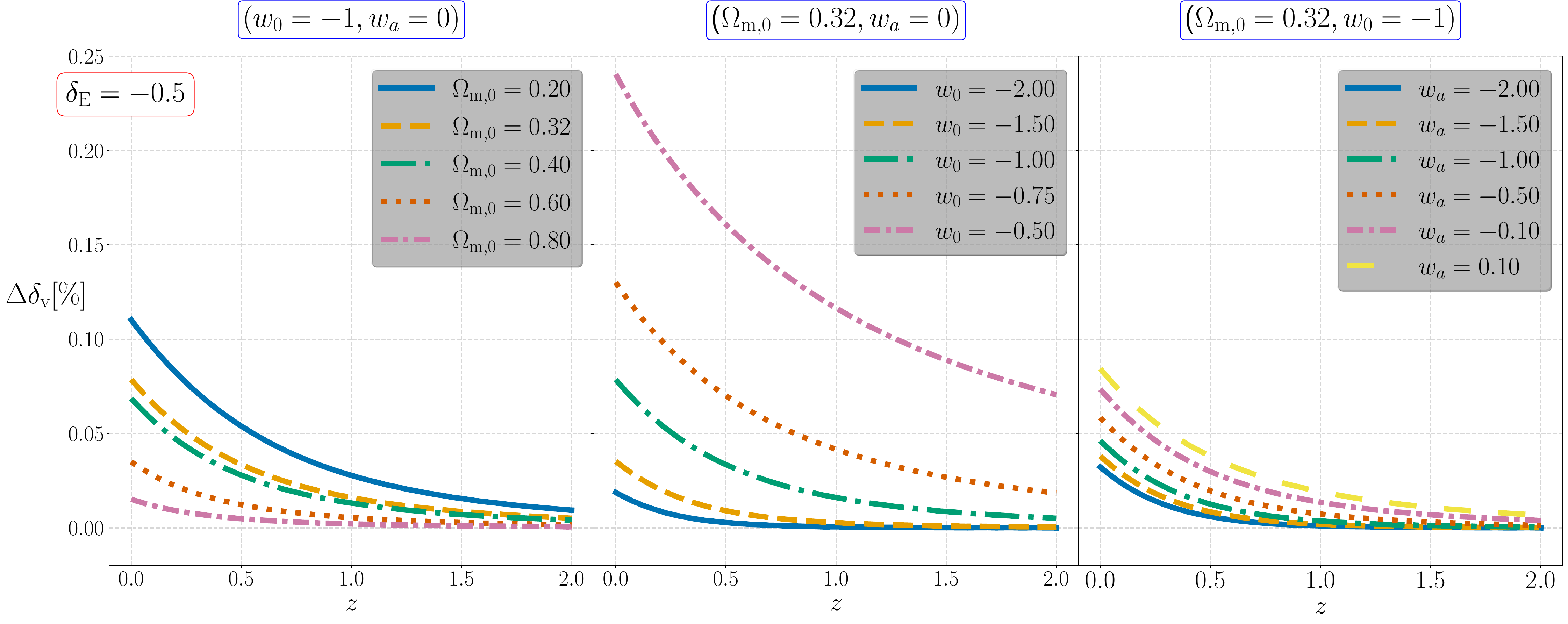}
    \caption{Impact of background cosmology on the mapping $\delta_{\rm v}(z, \delta_{\rm E} = -0.5)$, shown as a function of redshift in the range $z \in [0, 2]$ at fixed non-linear density contrast. Each column isolates the effect of a single cosmological parameter: $\Omega_{\mathrm{m},0}$ (left), $w_0$ (center), and $w_a$ (right), while keeping the others fixed.}
    \label{Fig:map_d}
\end{figure}
In figure~\ref{Fig:map_d}, we show the percentage difference in the mapping with respect to the EdS case (see eq.~\eqref{Eq:percentage_difference}) at fixed $\delta_{\rm E} = -0.5$, varying $\Omega_{\mathrm{m},0}$ (left), $w_0$ (center), and $w_a$ (right) within the ranges specified above. 

As shown in the plots, the percentage difference with respect to the EdS case is more pronounced at low redshift. This trend reflects the fact that, for a fixed non-linear density contrast $\delta_{\rm E}$, a void observed at higher redshift has evolved for a shorter time and has therefore been less affected by the presence of DE. In other words, the earlier the epoch, the closer the background evolution is to that of an Einstein–de Sitter universe, where structure formation is unimpeded by the accelerated expansion.

A second key feature is that the percentage difference is always positive, i.e.,
\begin{align}
    \delta_{\rm v}(z,\delta_{\rm E})_{\rm model} \,<\, \delta_{\rm v}(z,\delta_{\rm E})_{\rm EdS}\,,
\end{align}
meaning that, in models with DE, the linearly extrapolated density contrast corresponding to a given final (non-linear) void is more negative than in EdS.
This can be explained by considering two competing effects. On one hand, DE suppresses structure formation; as a result, reaching the same final underdensity ($\delta_{\rm E} = -0.5$ in this case) at redshift $z$ requires starting from a more negative initial underdensity $\delta_{\rm v,in}$ at $x_{\rm in}$. On the other hand, the growth factor in EdS is larger. However, this enhanced growth is not sufficient to compensate for the initially less extreme underdensity. The overall behavior results from a non-trivial combination of these two, with the dominant contribution coming from the modified initial conditions  as shown in the plots.  

Finally, we would like to discuss the dependence of the mapping on the underlying cosmology. Variations in $\Omega_{\mathrm{m},0}$, $w_0$, and $w_a$ effectively shift the redshift at which DE starts to dominate the expansion. For instance, increasing $\Omega_{\mathrm{m},0}$ delays the onset of DE domination, reducing its impact on void evolution and bringing the mapping closer to the EdS case. The same logic can be applied to $w_0$ and $w_a$: less negative values lead to an earlier onset of DE domination. This, in turn, enhances the deviation from the EdS mapping.
Among the three cosmological parameters considered, $w_0$ has the strongest impact, while variations in $w_a$ produce comparatively smaller changes. It is important to note, however, that the effect of $w_a$ becomes more relevant when $w_0$ is less negative than $-1$. In other words, models with $w_0 > -1$ exhibit a stronger response to variations in $w_a$. Although this specific dependence is not explicitly shown in the current plots, it is consistent with the behavior observed in the previous section (see figure~\ref{Fig:impact_w_a}). More generally, $w_0w_a$CDM models could induce larger deviations from the EdS mapping compared to the $\Lambda$CDM case. Although still small, such deviations may become relevant in the context of precision void analyses, as even small changes in the mapping could translate into potentially observable effects. In this regard, the use of a cosmology-dependent mapping may offer a useful tool for testing $w_0w_a$CDM models, especially when combined with other complementary observables.
\begin{figure}[t]  
\centering\includegraphics[width=1\textwidth]{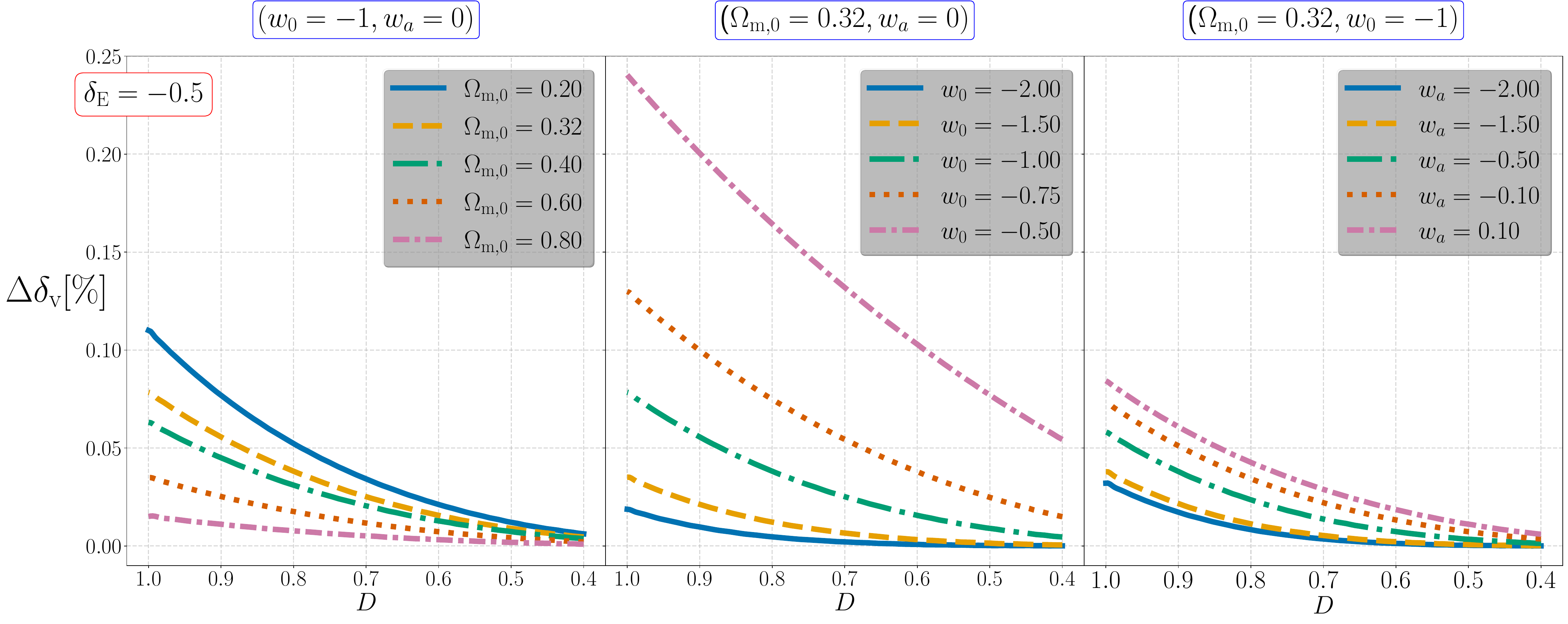}
    \caption{Impact of background cosmology on the mapping $\delta_{\rm v}(D,\delta_{\rm E}=-0.5)$. The mapping is displayed as a function of the linear growth factor $D$ over the interval $D\in[0.4,1]$. Each column isolates the effect of a single cosmological parameter, $\Omega_{\rm m,0}$ (left), $w_0$ (center), and $w_a$ (right), while the others are kept fixed.}
    \label{Fig:map_d_growth_factor}
\end{figure}

\begin{figure}[t]  
\centering\includegraphics[width=1\textwidth]{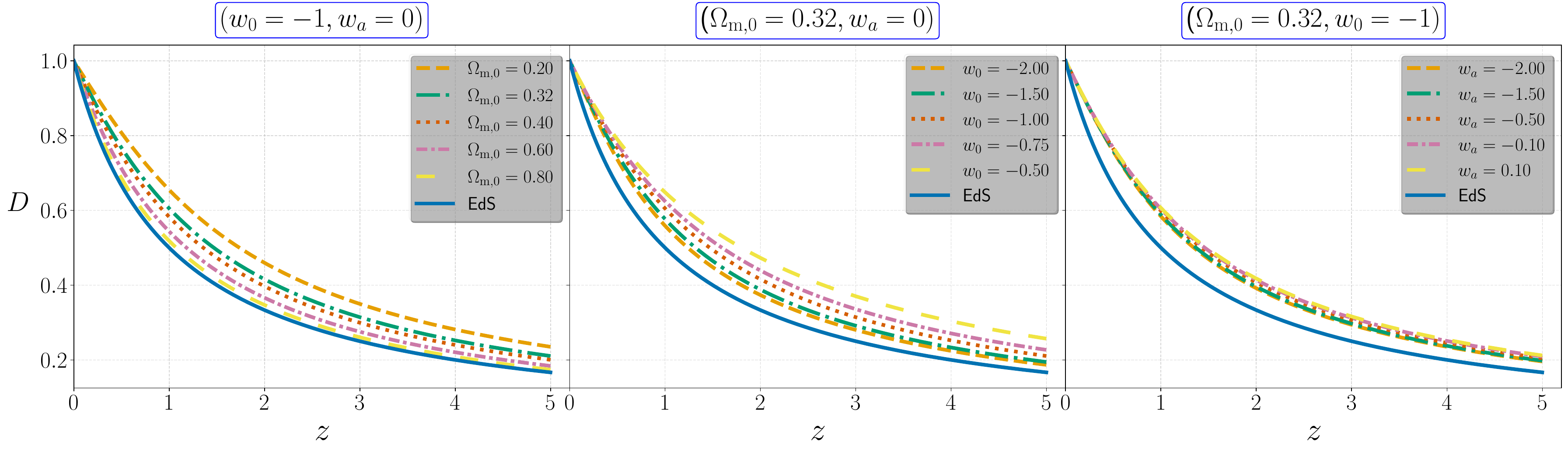}
    \caption{Linear growth factor $D(z)$ as a function of redshift in the range $z\in[0,5]$, normalized to unity today, $D(0)=1$. The three panels correspond to variations in $\Omega_{\rm m,0}$ (left), $w_0$ (center), and $w_a$ (right), while keeping the other parameters fixed.}
    \label{Fig:growth_factor}
\end{figure}

We complete the analysis by presenting in figure~\ref{Fig:map_d_growth_factor}  the cosmological dependence of the mapping as a function of the linear growth factor $D$, i.e. $\delta_{\rm v}(D,\delta_{\rm E})$. The results are presented at a fixed value of the non-linear density contrast, $\delta_{\rm E}=-0.5$, corresponding to voids that can be reliably identified in both observations and simulations.  The three columns explore the same cosmological variations as in the previous plots: $\Omega_{\rm m,0}$ (left), $w_0$ (center), and $w_a$ (right). In each panel, we display the percentage difference between a given cosmological model and the EdS case, namely
\begin{align}
    \Delta\delta_{\rm v}[\%](D,\delta_{\rm E})\,\equiv\,\frac{\delta_{\rm v}^{(\text{model})}(D,\delta_{\rm E}) - \delta_{\rm v}^{(\mathrm{EdS})}(D,\delta_{\rm E})}{\delta_{\rm v}^{(\mathrm{EdS})}(D,\delta_{\rm E})}\,\times\,100\,.
\label{Eq:percentage_difference_growth_factor}
\end{align}
We recall that the linear growth factor is defined as the solution of the linear equation, i.e. eq.~\eqref{Eq:linear_evolution_equation}, normalized to unity today, $D(z=0)=1$. To show its dependence on cosmology, we present in figure~\ref{Fig:growth_factor} the evolution of $D(z)$ over the redshift range $z\in[0,5]$, for the different $w_0w_a$CDM cosmological models with same variations in $\Omega_{\rm m,0}$ (left), $w_0$ (center), and $w_a$ (right) as in the previous plots. The interpretation of these trends is as follows: dark energy suppresses the (linear) growth of structures relative to the EdS case. Because of the normalization at $z=0$, we have that at fixed redshift, $D(z)$ is always larger in dark-energy models than in EdS. This does not signal enhanced growth, but simply reflects the fact that, with reduced growth efficiency (with respect to the EdS model), the solution must compensate at earlier times to reach the common endpoint $D(z=0)=1$.

Since $D(z)$ increases monotonically with decreasing redshift for any $w_0w_a$CDM cosmological model, as shown in figure~\ref{Fig:growth_factor}, one can always invert the relation to express $z(D)$ and use $D$ as a time variable. However, one must keep in mind that the relation between $D$ and $z$ is cosmology-dependent: at fixed $D$, each model reaches the given value for $D$ at different $z$. 

In the excursion-set formalism, and more generally, when studying the time evolution of the VSF, it is customary to use the growth factor as the evolution variable. The reason is that $D$ provides a ``natural way'' to connect the linear regime at early times with the linear regime at later times, allowing for the theoretical prediction (of the VSF) to be evolved forward or backward in time in a consistent manner, effectively providing a bidirectional link across cosmic epochs.
For this reason, we present in figure~\ref{Fig:map_d_growth_factor} the comparison between the mappings in EdS and $w_0w_a$CDM cosmologies as a function of the growth factor $D$, at a fixed non-linear density contrast $\delta_{\rm E}$. Moreover, this representation provides additional insight, as the comparison between EdS and $w_0w_a$CDM cosmologies is conceptually different when expressed as a function of the growth factor and when expressed as a function of the redshift (figure~\ref{Fig:map_d}). Indeed, for a given value of $D$, the corresponding redshift is always smaller in EdS than in a dark-energy cosmology, meaning that the comparison effectively traces voids of equal depth ($\delta_{\rm E} = -0.5$) at different cosmic epochs. Despite this conceptual difference, the results in figure~\ref{Fig:map_d_growth_factor} are identical to those in figure~\ref{Fig:map_d} once one maps $D$ (in figure~\ref{Fig:map_d_growth_factor}) back to $z$ using the relation of the dark-energy model under consideration. The reason is simple: the EdS mapping is redshift-independent, i.e.
\begin{align}
    \delta_{\rm v}^{\rm EdS}(z,\delta_{\rm E} = -0.5) \,=\, -0.86676\,.
\end{align}
So the percentage differences with respect to EdS remain unchanged whether one uses $z$ or the growth factor $D$. If, instead, one were to compare two dark-energy cosmologies, the outcome would differ. 

This result highlights a key point of the present analysis: expressing the mapping in terms of $D$ does not remove the differences introduced by a different cosmological background with respect to the EdS model. Although this expectation is sometimes stated in the literature, our results demonstrate that these differences persist. In fact, the deviations relative to EdS remain exactly the same, regardless of whether one uses redshift or the growth factor. These differences are physical, potentially observable, and cannot be absorbed by a redefinition of the time variable.

\begin{figure}[t]  
\centering\includegraphics[width=1\textwidth]{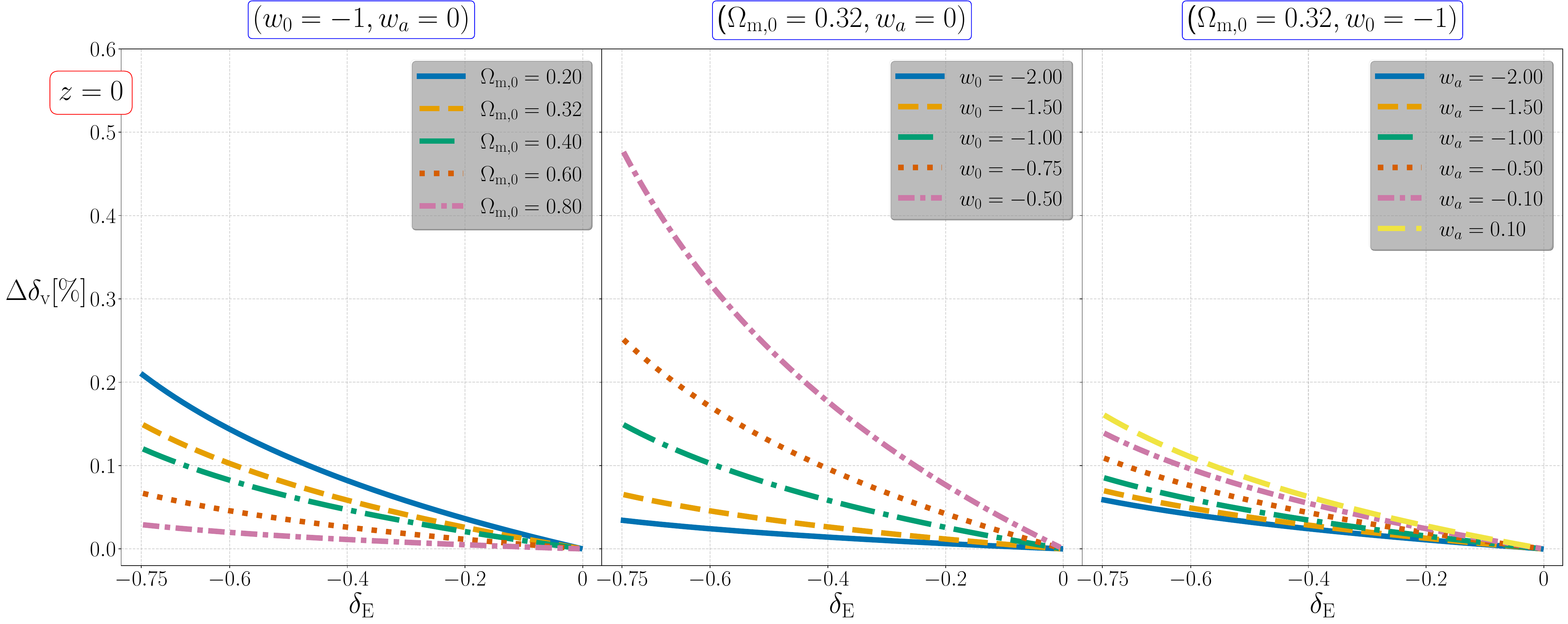}
    \caption{Impact of background cosmology on the mapping $\delta_{\rm v}(z=0,\delta_{\rm E})$, shown as a function
    of $\delta_{\rm E}$ in the range $\delta_{\rm E}\in[-0.75,\,-0.005]$ at $z=0$. Each column isolates the effect
    of a single cosmological parameter: $\Omega_{\rm m,0}$ (left), $w_0$ (center), and $w_a$ (right), while keeping the others
    fixed.}
    \label{Fig:map_z}
\end{figure}

Now, we present the final result of the mapping analysis in figure~\ref{Fig:map_z}. In this case, we fix the redshift at $z = 0$ and vary the non-linear matter density contrast $\delta_{\rm E}\in[-0.75,-0.005]$. Compared to the previous analysis, the deviations from the EdS case are larger. This indicates that the mapping $\delta_{\rm v}(z, \delta_{\rm E})$ is more sensitive to the depth of the void than to its redshift, at least within the ranges considered.
From figure~\ref{Fig:map_z}, we see that the strongest deviations occur for the deepest voids, i.e., for more negative values of $\delta_{\rm E}$. This trend reflects an increasing imbalance between the two effects previously discussed: the need for more negative initial conditions in DE models and the reduced linear growth relative to EdS. Since the linear growth factor is fixed for a given cosmology, ``its contribution remains unchanged'' across different values of $\delta_{\rm E}$. In contrast, the shift in the required initial conditions becomes progressively larger for deeper voids, as indicated by the plots. This growing asymmetry between the two effects results in increasingly stronger deviations from the EdS case as $\delta_{\rm E}$ decreases.
The dependence on cosmological parameters follows the same logic as before.

\subsection{Implementation of the shell-crossing}
\label{Sec:implementation_of_the_shell_crossing}
We now discuss the final case of interest in our analysis: the values of the linear and non-linear matter density contrast at the epoch of shell-crossing. Specifically, these values are determined using the condition derived here, i.e.~eq.~\eqref{Eq:Shell_crossing_Delta}, for evaluating the shell-crossing epoch within the hydrodynamical formalism.
The same computations can be carried out using the $R$-based formalism. However, since both approaches yield identical results, we present here only the case based on the hydrodynamical formalism. 
In appendix~\ref{Sec:Consistency_checks_with_the_EdS_model}, we have tested that the two conditions, i.e.~eqs.~\eqref{Eq:shell_crossing_condition_analytical} and \eqref{Eq:Shell_crossing_Delta}, allow us to accurately reproduce the known analytic result in the EdS case. The agreement holds within the numerical accuracy (see figure~\ref{Fig:comparison_shell_crossing_conditions}). This test is important both for validating the theoretical derivation of the new conditions and for demonstrating that either approach can be used interchangeably to compute the epoch of shell-crossing.

\begin{figure}[t]  
    \centering\includegraphics[width=1\textwidth]{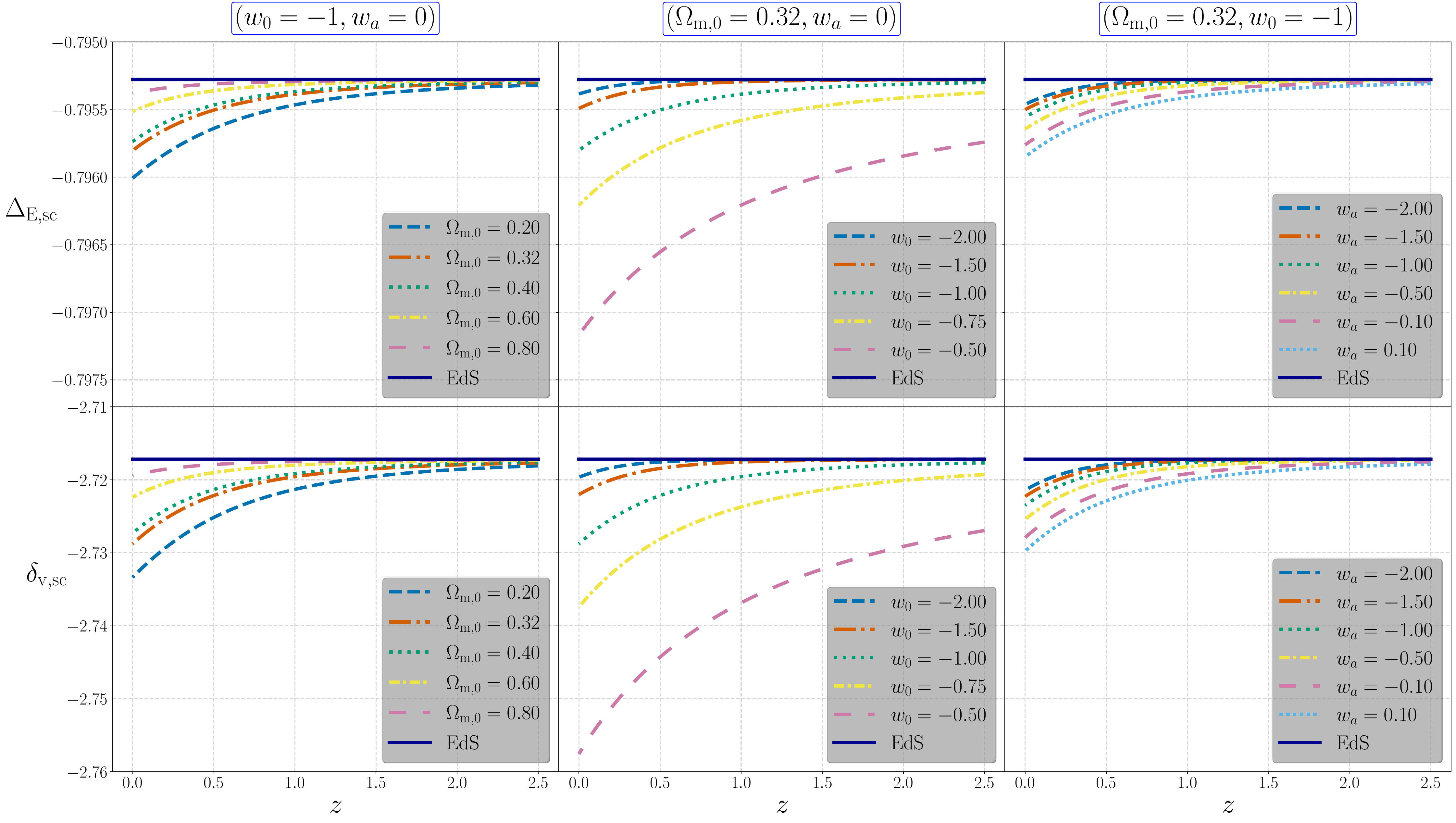}
    \caption{Top row: values of the non-linear matter density contrast at the moment of shell-crossing, $\delta_{\rm E,sc}$, as a function of redshift in the range $z \in [0, 2.5]$.
    Bottom row: corresponding linearly extrapolated values, $\delta_{\rm v,sc}$, evaluated at the same redshifts. Each column illustrates the impact of a single cosmological parameter, i.e.~$\Omega_{\mathrm{m},0}$ (left), $w_0$ (center), and $w_a$ (right).}
    \label{Fig:models_sc}
\end{figure}

We present our results in figure~\ref{Fig:models_sc}. The top row shows the values of the non-linear matter density contrast at the moment of shell-crossing, $\delta_{\rm E,sc}$, while the bottom row displays the corresponding linearly extrapolated values, defined as
\begin{align}
    \delta_{\rm v,sc}(z) \,=\, \delta_{\rm v}\big(z,\delta_{\rm E,sc}(z)\big)\,.
\end{align}
Before discussing the results, we briefly outline the numerical procedure used to generate the plots. The method closely follows the one described in section~\ref{Sec:impact_of_cosmology_on_single_void_evolution}, and is analogous to the standard approach used in the spherical collapse model to determine the collapse threshold $\delta_{\rm c}$ (see, e.g.,~\cite{Pace:2010sn}).

For each target redshift $z\in[0,2.5]$, we determine the initial conditions such that shell-crossing occurs precisely at that redshift.\footnote{We discretize the redshift interval uniformly into approximately fifty points. At each of these redshifts, we apply the procedure described below to compute $\delta_{\rm E,sc}$ and $\delta_{\rm v,sc}$. The final curves shown in the plots are obtained by interpolating the discrete results using a cubic spline.}
To this end, we integrate the non-linear evolution equation, eq.~\eqref{Eq:non_linear_evolution_equation}, using three ICs:
\begin{align}
    \delta_\mathrm{v,in_1} \,=\, \delta_{\rm v,in}\, [1 - \varepsilon]\,, \qquad
    \delta_{\rm v,in_2} \,=\, \delta_{\rm v,in}\,, \qquad
    \delta_{\rm v,in_3} \,=\, \delta_{\rm v,in}\, [1 + \varepsilon]\,,
\end{align}
with $\varepsilon = 10^{-4}$. At the final redshift, we evaluate the function
\begin{align}
    f(\delta_{\rm v,in}) = \frac{\mathrm{d}\delta_{\rm E}}{\mathrm{d}\delta_{\rm v,in}} + \frac{(1 + \delta_{\rm E})}{(1 + \delta_{\rm v,in})\,\delta_{\rm v,in}}\,,
\end{align}
where the derivative is computed numerically using the two-point centered difference method, accurate to order $\varepsilon^2$.

We then apply a shooting method to find the value of $\delta_{\rm v,in}$ satisfying $f(\delta_{\rm v,in}) = 0$, corresponding to the condition for shell-crossing, eq.~\eqref{Eq:Shell_crossing_Delta}. Once the correct initial condition has been identified, we integrate both the linear and non-linear equations up to the target redshift, obtaining  $\delta_{\rm E,sc}(z)$ and $\delta_{\rm v,sc}(z)$. For each redshift, we also check that the shell-crossing condition is not met at earlier times, verifying that $z$ corresponds to the first and only occurrence.

The three columns in figure~\ref{Fig:models_sc} correspond to variations in $\Omega_{\mathrm{m},0}$, $w_0$, and $w_a$, respectively, with the other parameters held fixed. The explored ranges are those introduced at the beginning of the section.
The overall behavior observed in the plots closely mirrors what was discussed in the previous section. First, the variations in the shell-crossing threshold across different cosmological models are small, typically below the percent level. This is analogous to what happens for the spherical collapse threshold $\delta_{\rm c}$, and implies that adopting a cosmology-dependent shell-crossing threshold is not expected to induce large deviations in void statistics—although percent-level effects remain theoretically relevant.
Second, all curves tend to converge to the EdS value at high redshift. This is consistent with the expectation that, as one moves to earlier epochs, the impact of DE becomes negligible and structure formation proceeds as in a matter-dominated universe. Conversely, at low redshift, DE becomes dynamically relevant, and the shell-crossing threshold deviates more significantly from the EdS case. This trend reflects the same physical mechanism discussed previously for the void mapping: the earlier DE starts to dominate, the more the void evolution is suppressed.
Finally, the dependence on cosmological parameters follows a consistent pattern. Variations in $\Omega_{\mathrm{m},0}$, $w_0$, and $w_a$ effectively shift the redshift at which DE becomes dominant. Increasing $\Omega_{\mathrm{m},0}$, or decreasing $w_0$, or $w_a$ delays the onset of DE domination, bringing the results closer to those of EdS. The interpretation is thus fully aligned with the discussion provided in the previous section.

\section{Conclusions}
\label{Sec:Conclusions}
In this work, we have developed a novel hydrodynamical framework to model the evolution of spherically symmetric cosmic voids. While similar techniques have long been applied to describe the spherical collapse of overdensities, this marks the first systematic application of the formalism to underdensities. Modeling matter as a pressureless fluid governed by the continuity, Euler, and Poisson equations in an expanding universe, offer a physically transparent and extensible approach to studying void formation on sub-horizon scales.

This hydrodynamical perspective offers several key advantages over the traditional \mbox{$R$-based} formulation. It provides a more intuitive interpretation of void dynamics, follows directly from the action—making it naturally compatible with extensions to modified gravity—and, crucially, enables a consistent and cosmology-dependent mapping between Lagrangian and Eulerian space. This latter result represents a significant step forward, as it allows for the first time a direct connection between linear theory and fully non-linear void observables in cosmologies beyond the EdS case.

We have applied this formalism to explore how the background cosmology influences the evolution of an isolated void. Focusing on $\Lambda$CDM and dynamical dark energy models such as $w_0$CDM and $w_0w_a$CDM, we have shown that variations in the present-day matter density $\Omega_{\mathrm{m},0}$ and the dark energy equation-of-state parameters $w_0$ and $w_a$ significantly affect the evolution of voids. These parameters enter the perturbation dynamics solely through their effect on the time-dependent matter fraction $\Omega_{\rm m}(t)$. Deviations as large as $30\%$ are observed in the explored parameter space, confirming that the evolution of the mean non-linear matter density contrast inside voids is highly sensitive to the dark energy equation of state. For shell-crossed voids, it is generally understood that a significant part of this effect arises from differences in linear growth. Crucially, we show that it is even more pronounced for the shallower, unshell-crossed voids that will be prime observational targets for surveys such as \textit{Euclid}, suggesting that this enhanced sensitivity could have a direct observational impact.
Interestingly, the largest deviations occur for parameter values compatible with current DESI constraints. If these measurements are confirmed, cosmic voids could provide an independent and powerful means of probing the nature of dark energy and the expansion history of the universe.

In addition, we have presented the first cosmology-dependent construction of the linear to non-linear mapping $\delta_{\rm v}(z, \delta_{\rm E})$ for voids, providing a numerical implementation of the transformation from Lagrangian to Eulerian space. Although the resulting corrections relative to the EdS case are typically at the sub-percent level, they could become relevant for high-precision void statistics in upcoming large-scale surveys, as it is the case for HMF. In particular, the ability to consistently connect initial underdensities to late-time Eulerian observables opens the door to more accurate modeling of void bias, void size functions, and their sensitivity to cosmological parameters—further enhancing the role of voids as cosmological probes.

Another major theoretical result is our derivation of new exact conditions for shell-crossing in general cosmologies. Using both the $R$-based and hydrodynamical approaches, we have identified the shell-crossing epoch by dynamically determining the critical Eulerian density $\delta_{\rm E,sc}$ that marks the onset of multistreaming. This represents the first fully consistent implementation of such a condition in the literature. We have computed this threshold across different dark energy scenarios, finding that deviations from the EdS result remain small but potentially significant for precision modeling.

All results in this work have been obtained through numerical integration, and we plan to publicly release the code developed for this study. This tool will enable the broader community to explore the evolution of voids across a wide range of cosmological models and to exploit voids as precise and complementary probes of the dark sector.

\acknowledgments
We are grateful to the referee for their valuable comments and insights, which have helped improve the manuscript.
We thank C. Carbone for valuable discussions during the initial stages of this work.
T.M. wishes to express his gratitude to Philip Mocz and Jeff Jennings  for helpful discussions that contributed to improving the numerical code.
T.M. acknowledges Emiliano Sefusatti, Mauro Moretti, Baojiu Li, Francesco Giuseppe Capone, Andrea Ponticelli, and Pietro Pellecchia for their valuable input and insightful comments.
T.M., N.F. and F.P. acknowledge the Fundação para a Ciência e a Tecnologia (FCT) project with ref. number PTDC/FIS-AST/0054/2021 and the COST Action CosmoVerse, CA21136, supported by COST (European Cooperation in Science and Technology). T.M. and N.F. acknowledge the Istituto Nazionale di Fisica Nucleare (INFN) Sez. di Napoli, Iniziativa Specifica InDark.
T.M. and G.V. acknowledge support from the Simons Foundation to the Center for Computational Astrophysics at the Flatiron Institute.
F.P. acknowledges partial support from the INFN grant InDark and from the Italian Ministry of University and Research (\textsc{mur}), PRIN 2022 `EXSKALIBUR – Euclid-Cross-SKA: Likelihood Inference Building for universe's Research', Grant No.\ 20222BBYB9, CUP C53D2300131 0006, and from the European Union -- Next Generation EU.

\appendix

\section{Void formation in EdS universe}
\label{Sec:Appendix_EdS}
In this appendix, we review the spherical model for void formation in an EdS universe (see~\cite{Sheth:2003py,Blumenthal:1992ert,Lilje:1991oiu}).  
Integrating eq.~\eqref{Eq:Newton_equation} with respect to the cosmic time, we get
\begin{align}
    \frac{ \dot{R}^2}{2} - \frac{GM(R)}{R} \,=\, C\,,
    \label{Eq:EdS_energy_conservation}
\end{align}
where $C$ is a function of $r_{\rm in}$, and we recall that $R(t,r_{\rm in})$ denotes the radius of each shell (see the discussion in section~\ref{Sec:the_spherical_model}). Using the definition of the mass in eq.~\eqref{Eq:mass_conservation}, we can write
\begin{align}
    \dot{R}^2 -  H^2 R^2 \left[1 + \Delta_{\rm E}(R,t)\right] &=  \, 2C\,,
    \label{Eq:Integration_Newton_2}
\end{align}
where $\Delta_{\rm E}(R,t)$ is the mean density contrast defined in eq.~\eqref{Eq:mean_density_contrast} and $H$ is the Hubble parameter. To determine $C$ in eq.~\eqref{Eq:Integration_Newton_2} we evaluate the left-hand side at the initial time $t_\mathrm{in}$, where the perturbations are still linear. The initial velocity $\dot{R}(t_{\rm in},r_{\rm in})$ profile is determined by using the linear solution for the peculiar
velocity profile in a matter-dominated universe (see eq.~(5.119) of~\cite{Peebles:1980yev})
\begin{align}
    \dot{R}(t_{\rm in},r_{\rm in}) \,=\, H_{\rm in}r_{\rm in}  \left[1- \frac{1}{3}  \Delta_{\rm in}(r_{\rm in})\right]\,,
\end{align}
where $\Delta_{\rm in}(r_{\rm in}) \equiv \Delta_{\rm E}(r_{\rm in},t_{\rm in})$ and we recall that the subscript ``in'' refers to quantities evaluated at the initial time $t_{\rm in}$.
Thus, keeping only the first-order contribution in $\Delta_{\rm in}$, we get
\begin{align}
    C(r_{\rm in}) \,=\, -\frac{5}{6}\left( r_{\rm in} H_{\rm in} \right)^2   \Delta_{\rm in}\,.
    \label{Eq:EdS_constant}
\end{align}
Using eq.~\eqref{Eq:EdS_constant} and the mass conservation equation, we rewrite eq.~\eqref{Eq:EdS_energy_conservation} as
\begin{align}
    \frac{\dot{R}^2}{R^2} \,=\, H_{\rm in}^2 \left[ -\frac{5}{3}\left( \frac{r_{\rm in}}{R} \right)^2   \Delta_{\rm in} + \left( \frac{r_{\rm in}}{R} \right)^3 (1 + \Delta_{\rm in}) \right]\,. 
    \label{Eq:Newton_Equation_EdS}
\end{align}
The dynamical variable is the ratio $R/r_{\rm in}$, implying that the solution is independent of the initial radius. 
The solution for each shell in eq.~\eqref{Eq:Newton_Equation_EdS} can be derived in a parametric way, defining 
\begin{align}
    p(r_{\rm in}) \,=\, \frac{\dot{R}}{r_{\rm in}}\,\frac{1}{\sqrt{-\frac{5}{3}\Delta_{\rm in}H_{\rm in}^2}}\,.
    \label{Eq:def_p}
\end{align}
Then, eq.~\eqref{Eq:Newton_Equation_EdS} and the derivative of eq.~\eqref{Eq:def_p} with respect to time provide the following system
\begin{align}
    \frac{R}{r_{\rm in}} &= \frac{\left(1+\Delta_{\rm in}\right)\left(-\frac{5}{3}\Delta_{\rm in}\right)^{-1}}{p^2-1}\, ,\qquad\qquad
    \mathrm{d}t = -\left[\frac{2\left(1+\Delta_{\rm in}\right)}{H_{\rm in}\left(-\frac{5}{3}\Delta_{\rm in}\right)^{\frac{3}{2}}}\right]\,\frac{\mathrm{d}p}{\left(p^2-1\right)^2}\,.
    \label{Eq:system_EdS}
\end{align}
Changing variable to
\begin{align}
    \frac{1}{p} = \tanh{\frac{\left(\Theta+\Theta_{\rm in}\right)}{2}}\,,
\end{align}
and imposing the initial condition to determine $\Theta_{\rm in}$, we get
\begin{align}
    \frac{R(\Theta)}{r_{\rm in}} &= \frac{1 + \Delta_{\rm in}}{2}\left(-\frac{5}{3}\Delta_{\rm in}\right)^{-1} \left[\cosh \left(\Theta+\Theta_{\rm in}\right) -1\right]\,, \label{Eq:R_EdS(Theta)}\\
    \frac{t(\Theta)}{t_\mathrm{in}} &= \frac{3}{4}\left(1+\Delta_{\rm in}\right)\left(-\frac{5}{3}\Delta_{\rm in}\right)^{-\frac{3}{2}}\left[\sinh \left(\Theta+\Theta_{\rm in}\right) -\left(\Theta+\Theta_{\rm in}\right)\right]\,,
    \label{Eq:EdS:t(theta)}
\end{align}
where $\Theta\geq0$ is the parameter with respect to which we solve the system and $\Theta_{\rm in} = 2\sqrt{-\frac{5}{3}\Delta_{\rm in}}$. Using the mass conservation equation and that, in an EdS universe, 
$ H \, = \, 2/3t$,
we can derive the evolution of the mean density contrast as
\begin{align}
    1 + \Delta_{\rm E}(R,t) \,=\, \frac{9}{2}\,\frac{\left[\sinh\left(\Theta+\Theta_{\rm in}\right) - \left(\Theta+\Theta_{\rm in}\right)\right]^2}{\left[\cosh\left(\Theta+\Theta_{\rm in}\right) - 1\right]^3}\,.
\end{align}
In the literature, the parameter $\Theta_{\rm in}$ is usually set to zero even if in this way a singular behaviour in $\Theta = 0$ is introduced. Generally, this is used because it does not affect the calculation of the shell-crossing epoch. 

Having derived the general solutions for all shells, we now specialize them to the case of an inverse top-hat density profile, as defined in eq.~\eqref{Eq:initial_top_hat_delta}. In this case, it is sufficient to follow the evolution of the outermost shell, since all shells with $r_{\rm in} < r_{\rm v,in}$ (where $r_{\rm v,in}$ denotes the initial void radius) evolve identically due to the uniform density within each sphere. The solutions are then given by eqs.~\eqref{Eq:R_EdS(Theta)} and~\eqref{Eq:EdS:t(theta)}, upon making the substitutions $r_{\rm in} \to r_{\rm v,in}$ and $\Delta_{\rm in}(r_{\rm in}) \to \delta_{\rm v,in}$.

We now discuss the derivation of the \textit{epoch of shell-crossing} (see section~\ref{Sec:Shell_Crossing}), which we define as the moment when the outermost shell of the void intersects the shells of the background environment. As emphasized in section~\ref{Sec:Shell_Crossing}, although we adopt an idealized inverse top-hat profile for our calculations, it is conceptually useful to keep in mind the more realistic picture introduced there, based on a smoothed step-like density distribution (see figure~\ref{Fig:step_function}). In that framework, shell-crossing occurs progressively, starting near the void boundary. However, when adopting a top-hat profile, the only meaningful event is the final crossing of the outermost void shell with the background. 

To determine the moment of shell-crossing, we use the condition given in eq.~\eqref{Eq:shell_crossing_condition_analytical}, without repeating the theoretical discussion already provided earlier. So differentiating the two parametric solutions, taking $\mathrm{d}t = \mathrm{d}R = 0$, and working with $\Delta_{\rm in} \ll 1$, we get
\begin{align}
    0 &=\left[\cosh{\left(\Theta+\Theta_{\rm in}\right)}-1\right]\left(\frac{\mathrm{d}r_{\rm in}}{r_{\rm in}} -  \frac{\mathrm{d}\Delta_{\rm in}}{\Delta_{\rm in}}\right) +  \sinh{\left(\Theta+\Theta_{\rm in}\right)}\left[\mathrm{d}\Theta -\frac{5}{3}\left(-\frac{5}{3}\Delta_{\rm in}\right)^{-\frac{1}{2}}\mathrm{d}\Delta_{\rm in}\right],\\
    0 &= -\frac{3}{2}\frac{\mathrm{d}\Delta_{\rm in}}{\Delta_{\rm in}}\left[\sinh\left(\Theta+\Theta_{\rm in}\right)-\left(\Theta+\Theta_{\rm in}\right)\right] + \left[\cosh{\left(\Theta+\Theta_{\rm in}\right)}-1\right]\left[\mathrm{d}\Theta -\frac{5}{3}\left(-\frac{5}{3}\Delta_{\rm in}\right)^{-\frac{1}{2}}\mathrm{d}\Delta_{\rm in}\right]\,.
\end{align}
Although the above expressions seem to involve three independent differentials, i.e.~$\mathrm{d}r_{\rm in}$, $\mathrm{d}\Delta_{\rm in}$, and $\mathrm{d}\Theta$, we stress—as discussed in the main text (see section~\ref{Sec:Shell_Crossing})—that the problem depends on only two independent variables. In the main text, these are $t$ and $r_{\rm in}$, while here we have traded time $t$ for the parametric variable $\Theta$. Since the only independent variables are $r_{\rm in}$ and $\Theta$, and the two conditions above provide independent constraints, the system is well posed and can be solved consistently. 

Strictly speaking, $r_{\rm in}$ is the natural independent variable of the problem; in the case of an inverse top-hat profile, we are only interested in the behavior near the boundary of the void. Within this narrow region, the mapping between $r_{\rm in}$ and $\Delta_{\rm in}$ is monotonic (see eq.~\eqref{Eq:initial_top_hat_delta}), allowing us to equivalently express the dynamics in terms of either variable. While this observation may appear trivial, it will be useful in the discussion that follows. We can explicitly disregard the inner region $r_{\rm in} < r_{\rm v,in}$, as no shell-crossing can occur there due to the uniform density.

Thus, to determine the moment of shell-crossing, it is sufficient to express the density differential in terms of the radial coordinate—or vice versa—depending on which is more convenient, as the two are not independent near the boundary.  However, as we are considering a top-hat profile, the radial derivative of the density is discontinuous at the boundary between the void and the surrounding environment, i.e.
\begin{align}
    \frac{\mathrm{d} \ln \Delta_\mathrm{in}}{\mathrm{d} \ln r_{\rm in}} = 
    \begin{cases}
        \,0 & \text{for } r_{\rm in} < r_{\rm v,in} \\
        \,-3 & \text{for } r_{\rm in} > r_{\rm v,in}
    \end{cases}\,.
\end{align}
As in section~\ref{Sec:Shell_Crossing}, we must therefore distinguish whether the radial derivative of the density is taken from the inside or from the outside, and now examine the consequences of these two choices in the EdS case.

\paragraph{The left-hand case. }
Taking the left-hand derivative and expressing $\mathrm{d}\Delta_{\rm in}$ as a function of $\mathrm{d}r_{\rm in}$, we obtain
\begin{align}
    \begin{pmatrix}
        \left[\cosh{\left(\Theta+\Theta_{\rm in}\right)}-1\right] & \sinh{\left(\Theta+\Theta_{\rm in}\right)}\\
        0 & \left[\cosh\left(\Theta+\Theta_{\rm in}\right)-1\right]
    \end{pmatrix}
    \begin{pmatrix}
        \frac{\mathrm{d}r_{\rm in}}{r_{\rm in}} \\ \mathrm{d}\Theta
    \end{pmatrix}
    \, = \,
    \begin{pmatrix}
        0 \\ 0
    \end{pmatrix}\,,
\end{align}
where we have decided to take $\mathrm{d}r_{\rm in}/r_{\rm in}$ as a variable for dimensional reasons.
Setting the determinant to zero, we obtain $\Theta + \Theta_{\rm in} = 0$. Since $\Theta_{\rm in} > 0$ and $\Theta \geq 0$, this equation admits no solution, indicating that no shell-crossing occurs. This is consistent with the discussion in section~\ref{Sec:Shell_Crossing}, as taking the left-hand derivative is physically equivalent to asking whether two shells in a homogeneous and isotropic universe cross. This equivalence arises because all shells with $r < r_{\rm v,in}$ share the same density contrast in the case of an inverse top-hat profile, and thus evolve identically without ever intersecting.

We also note that, had we not imposed the initial conditions properly to determine $\Theta_{\rm in}$, the solution $\Theta = 0$ would have formally satisfied the equation. However, this would not correspond to a genuine shell-crossing event, but rather to a spurious solution resulting from an incorrect treatment of the initial conditions.

\paragraph{The right-hand case. }
In this case, we express $\mathrm{d}r_{\rm in}$ as a function of $\mathrm{d}\Delta_{\rm in}$ for computational convenience, and obtain
\begin{align}
\begin{pmatrix}
    -\frac{4}{3}\left(\cosh{Z}-1\right)  +  \sinh{Z}\left(-\frac{5}{3}\Delta_{\rm in}\right)^{\frac{1}{2}} &
    \sinh{Z}\\
     -\frac{3}{2}\left[\sinh{Z}-Z\right]+\left[\cosh{Z}-1\right]\left(-\frac{5}{3}\Delta_{\rm in}\right)^{\frac{1}{2}} & \cosh{Z}-1
\end{pmatrix}
\begin{pmatrix}
     \frac{\mathrm{d}\Delta_{\rm in}}{\Delta_{\rm in}} \\
     \mathrm{d}\Theta
\end{pmatrix}
%
= \begin{pmatrix}
    0 \\ 0
\end{pmatrix}\,,
\end{align}
where for simplicity we have defined $Z = \left(\Theta+\Theta_{\rm in}\right)$. Imposing the determinant of the matrix to be zero, we get
\begin{align}
    \frac{\sinh{Z}\left(\sinh{Z}-Z\right)}{\left(\cosh{Z}-Z\right)^2} \, = \, \frac{8}{9}\,,
\end{align}
which can be solved numerically 
\begin{align}
    Z_{\mathrm{sc}} = \left(\Theta+\Theta_{\rm in}\right)_{\mathrm{sc}} \approx \Theta_{\mathrm{sc}} \, = 3.48752\,,
\end{align}
where the subscript ``sc'' stands for shell-crossing.
Now, we are ready to determine $\delta_{\rm v,sc}$. To do this, we use that in an EdS universe the linear evolution equation can be solved to give
\begin{align}
    \delta_{\rm L}(t) \, = \, \delta_{\rm v,in}\left(\frac{t}{t_\mathrm{in}}\right)^{\frac{2}{3}}\,.
\end{align}
Thus, using eq.~\eqref{Eq:EdS:t(theta)} we get
\begin{align}
    \delta_{\rm v,sc} \, \approx \, -\frac{3}{5}\left\{\frac{3}{4}\left(1+\Delta_{\rm in}\right)\left[\sinh\Theta_\mathrm{sc} -\Theta_\mathrm{sc}\right]\right\}^{\frac{2}{3}}\, \approx -2.71718\,,
    \label{Eq:EdS_shell_crossing_value}
\end{align}
where we have used that $\Delta_{\rm in} \,\ll\, 1.$ The result in eq.~\eqref{Eq:EdS_shell_crossing_value} is redshift independent, in the sense that it does not depend on the redshift at which the void undergoes shell-crossing.

\section{Assumptions and setup for numerical integration}
\label{Sec:the_numerical_integration}
This section outlines the procedure to set the ICs for the numerical integration of eqs.~\eqref{Eq:Newton_equation} and \eqref{Eq:non_linear_evolution_equation} in the case the initial matter density profile is an inverse top-hat as in eq.~\eqref{Eq:initial_top_hat_delta}.
The ICs procedure is described in section~\ref{Sec:the_initial_conditions}, with additional issues addressed in sections~\ref{Sec:the_linear_regime} and~\ref{Sec:The_decaying_mode}.
\subsection{The initial conditions}
\label{Sec:the_initial_conditions}
We now outline the standard procedure to set ICs for the hydrodynamical formalism, and subsequently extend it to the $R$-based framework.

Since eq.~\eqref{Eq:non_linear_evolution_equation} is a second-order differential equation, two conditions are necessary. 
The first one is set at \textit{early time} by making the following two assumptions:
\begin{enumerate}
    \item Matter perturbations are linear at $t_\mathrm{in}$, i.e., $\delta_\mathrm{L} = \delta_{\rm E}$, during matter domination.
    This yields the well-known general solution at early times~\cite{Peebles:1980yev}:
    \begin{align}
        \delta_{\rm E}(t) \,=\, \delta_{\rm L}(t) \,=\, A
        \,\exp{\left(x-x_{\rm in}\right)}
        + B
        \,\exp{\left[-\frac{3}{2}\left(x-x_{\rm in}\right)\right]}\,,
        \label{Eq:matter_dominance_solution}
    \end{align}
    where $x_{\rm in} = \ln a_{\rm in}$, and $A$, $B$ are real constants. The first term represents the growing mode, and the second the decaying mode. 
    \item  As standard procedure, the decaying mode is assumed to be negligible, i.e.~$B = 0$.
\end{enumerate}
Thus, combining these two assumptions, one can fix the ICs for eq.~\eqref{Eq:non_linear_evolution_equation} as
\begin{align}
        \delta_{\rm E}(x_{\rm in}) = 
        \delta^{'}_{\rm E}(x_{\rm in}) = \delta_{\rm v,in}\,.
    \label{Eq:the_initial_conditions_delta}
\end{align}
The initial value for the matter density contrast $\delta_{\rm v,in}$ is not fixed a priori but determined by imposing a \textit{late-time} condition on the non-linear matter density contrast; this is the second condition we impose to solve eq.~\eqref{Eq:non_linear_evolution_equation}. For instance, to model a void with $\delta_{\rm E} = -0.5$ at $z = 0$, we iterate on $\delta_{\rm v,in}$ until the solution satisfies this late-time condition. This defines our numerical procedure for solving eq.~\eqref{Eq:non_linear_evolution_equation}.

In the $R$-based approach, one has to solve eq.~\eqref{Eq:Newton_equation}, for which we specify the initial radius and its time derivative. Owing to the homogeneity of the equation in 
$R$, the initial radius is arbitrary, while the time derivative is set by using the linear solution for the peculiar velocity profile in a matter-dominated universe (see eq.~(5.119) of~\cite{Peebles:1980yev}) under the assumption of spherical symmetry. We get 
\begin{align}
    \frac{R^{'}(x_{\rm in},r_{\rm v,in})}{R(x_{\rm in},r_{\rm v,in})} = \left( 1 -\frac{\delta_{\rm v,in}}{3}\right)\,.
\end{align}
Thus, we can proceed exactly as previously discussed, using an iterative procedure on $\delta_{\rm v,in}$. We now turn to a discussion of the two hypotheses introduced in this section.
\subsection{The linear regime}
\label{Sec:the_linear_regime}
Let us start by discussing the assumption that at $t_\mathrm{in}$, the matter perturbations are linear. Although it is \textit{qualitatively} understood that perturbations at $z\sim 1100$ are approximately linear, a more precise \textit{quantitative} criterion is necessary for computational purposes. Specifically, it is essential to determine the time at which the linear approximation ceases to be valid. One possible approach is to define this threshold as the time when the relative difference between the linear and non-linear density contrasts exceeds a chosen tolerance, such as 
0.1\%. Although this criterion is somewhat arbitrary, it provides a practical benchmark for assessing the limits of linear theory in the context of the desired numerical accuracy.

In our case, this problem manifests itself in a non-intuitive way. We fix an initial integration point $x_{\rm in}$, assume that the equations are linear at that time, perform the integration, and impose a late-time condition for the non-linear density contrast.
However, it is not guaranteed that this approach will achieve the desired level of precision at late times because we do not know a priori how accurate the assumption of linearity is at $x_{\rm in}$. Thus, what we observe is that the linear solution is not stable when the initial integration point is changed.\footnote{In this context, by ``linear solution'' we refer to the numerical integration of eq.~\eqref{Eq:linear_evolution_equation}, solved using the same ICs, given in eq.~\eqref{Eq:the_initial_conditions_delta}, for the non-linear one.} The value of the linear solution at late times changes when the initial integration point is modified. However, the non-linear solution remains stable, as its value is fixed at late times. 
\begin{figure}[t!]
    \centering
    \includegraphics[width=1.0\textwidth]{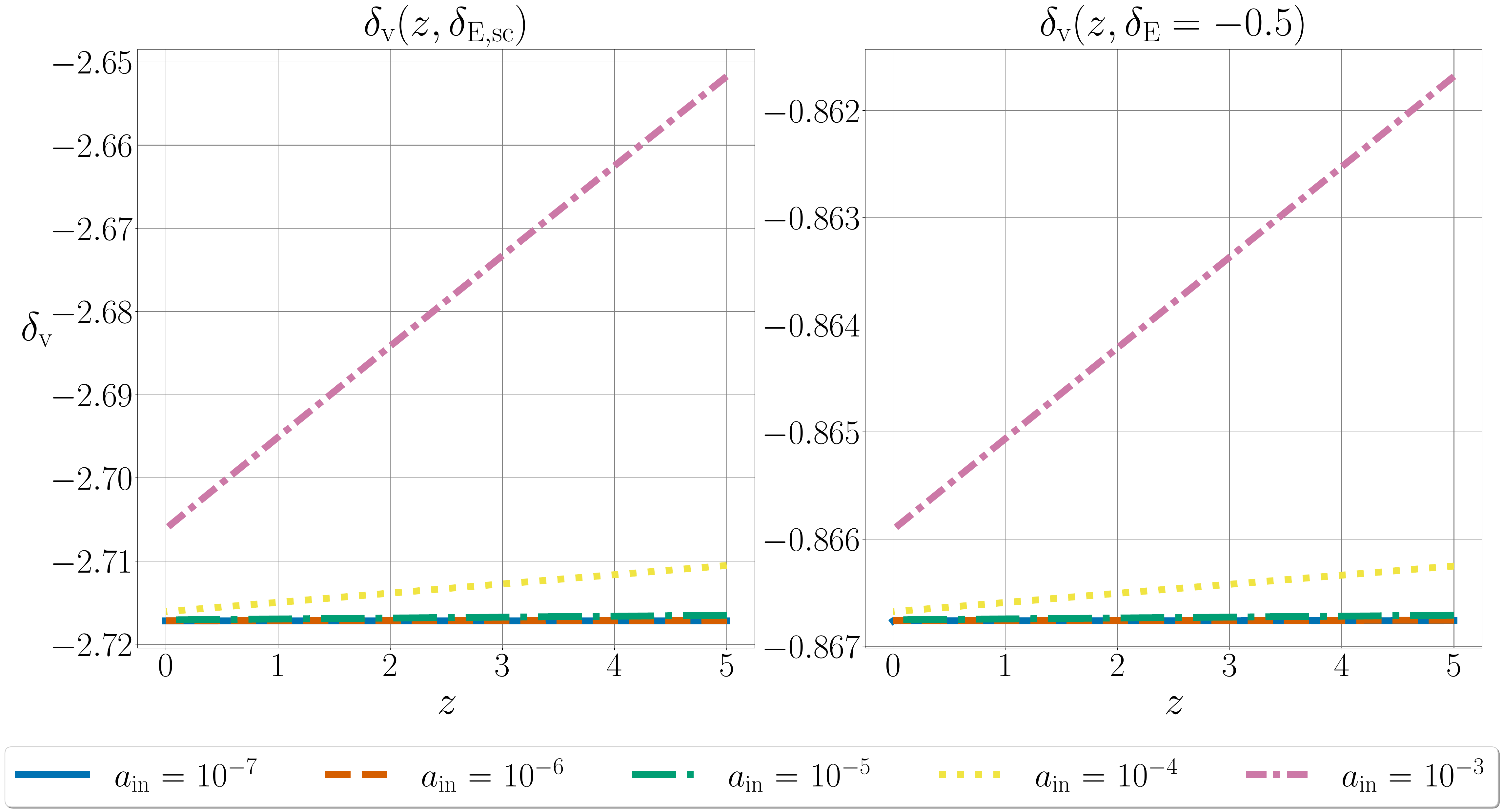}
    \caption{
    Left panel: dependence of the linearly extrapolated matter density contrast $\delta_{\rm v}$ at shell-crossing $\delta_{\rm E,sc}$ on the initial time of the integration $a_\mathrm{in}$. Right panel: dependence of the linearly extrapolated matter density contrast $\delta_{\rm v}$ at $\delta_{\rm E}=-0.5$ on the initial time of the integration $a_\mathrm{in}$.}
    \label{Fig:init_condition}
\end{figure}
We show two concrete cases in figure~\ref{Fig:init_condition}, both within an EdS universe:
\begin{itemize}
    \item Case 1 is shown in the left panel of figure~\ref{Fig:init_condition}. We study the evolution of voids by imposing that they reach the epoch of shell-crossing (see section~\ref{Sec:Shell_Crossing}), i.e.~$\delta_{\rm E} = \delta_\mathrm{E,sc}$, at a specific redshift $\bar{z}\in[0,5]$. At $\bar{z}$, we evaluate the linearly extrapolated density contrast $\delta_{\rm v}(\bar{z},\delta_{\rm E,sc})$. The plot shows how $\delta_{\rm v}$ changes with redshift, although in an EdS universe it is expected to remain constant.
    \item Case 2 is shown in the right panel of figure~\ref{Fig:init_condition}. We fix the non-linear density contrast to $\delta_{\rm E} = -0.5$ at different times within the same redshift range, and then compute the corresponding linearly extrapolated value. 
\end{itemize}

As we can see in both cases, by pushing the initial point backward in time, we find that the results reproduce those of the EdS model and remain stable. This behaviour is not related to the underlying physics; rather, by moving the initial point backward in time, we ensure that we start integrating the equations deeply within the linear regime. This is an artifact that arises from our lack of precise knowledge regarding the values of the linear and non-linear density contrasts at the initial time.
Thus, to establish the initial integration point for the equations, we choose the point for which the results for $\delta_{\rm v}$ in the range $z\in(0,5)$\footnote{This is approximately the redshift range of LSS surveys.} converge to the EdS value and remain stable up to the fifth decimal place. This corresponds to an accuracy of approximately $0.005\%$ with respect to the analytical EdS solution. In the present case, we set
\begin{align}
    a_\mathrm{in} = 10^{-7}\,, \quad x_\mathrm{in} \approx-16.12 \,,\quad z_\mathrm{in} \approx10^{7} \,.
    \label{Eq:value_x_in_z_in}
\end{align}
All points for which $a_\mathrm{in} < 10^{-7}$ are also found to yield stable performance. However, the farther we go backward in time, the slower our numerical solver becomes if we want to maintain the same precision. Therefore, the selected parameter value represents an optimal balance between computational speed and solution accuracy.

It is essential to emphasize that our methodology assumes that the radiation contribution is precisely zero at both the background and perturbative levels. If radiation is present at the perturbation level, the master equation is no longer valid. If it is included at the background level, the explicit form of the growing and decaying modes differs from that shown in eq.~\eqref{Eq:matter_dominance_solution} (see~\cite{Padmanabhan:1996qwe}), and it is no longer true that going back in time improves the validity of the linear regime.
This has to be taken into account for studying the evolution of voids at redshifts where the contribution of radiation cannot be neglected. However, for voids targeted by ongoing and upcoming galaxy surveys, the radiation contribution is negligible. Indeed, this model accurately captures the evolution of realistic voids in the redshift range $z \in (0,5)$, which corresponds to the epoch of structure formation.
Therefore, pushing $t_\mathrm{in}$ back in time should be seen solely as an artifact introduced to improve the precision of the numerical solution. It neither alters the physical assumptions of the model nor requires the inclusion of radiation. 

We would like to conclude this section by emphasizing that this level of precision—the accuracy of the linear and non-linear solutions at the fifth decimal place—is completely irrelevant when comparing the theoretical prediction for void statistics with numerical simulations or real data. In this context, there are other effects that have an impact on the results and their interpretation; for example, the fact that real voids are not perfectly spherical.  
The point of the present discussion is purely theoretical and aims at understanding the origin of the observed dependence on $x_\mathrm{in}$. This is not a numerical artifact, but rather stems from the assumptions made in solving the equations themselves. If it were a numerical artifact, such as those that arise from varying the integration step, we would expect random behaviour, not a linear dependence like the one shown in figure~\ref{Fig:init_condition}. 

Finally, we stress that the same issue arises in the context of the spherical collapse model (see~\cite{Pace:2017qxv}), where the collapse threshold $\delta_{\rm c}$ exhibits a dependence on the initial integration time. The solution to this problem is exactly the same as the one presented in this section; see~\cite{Pace:2017qxv} for further details.

\subsection{The decaying mode}
\label{Sec:The_decaying_mode}
Another important aspect to consider is the assumption of neglecting the decaying mode, which is commonly adopted as a working hypothesis in this type of computation.  
Here, we provide a justification for this assumption and assess its range of validity in the model discussed in this paper.

\textit{Observationally}, this is justified by the fact that we see that structures grow over time. If the decaying mode had any significant impact, we would not observe the consistent growth of structures. A dominant decaying mode would suppress all perturbations, preventing the formation of structures. The coexistence of both modes would instead lead to a ``mixed'' evolution which, however, is not what we observe.

\textit{Theoretically}, this can be justified within a complete $\Lambda$CDM (or $w_0w_a$CDM) framework, where perturbations are generated during inflation. This is done by matching the super-horizon solutions for matter perturbations in Fourier space with the sub-horizon ones. In this way, it is possible to set the amplitude ($A$ and $B$ in the language of eq.~\eqref{Eq:matter_dominance_solution}) of the growing and decaying modes at horizon entrance~\cite{Padmanabhan:1996qwe}. If the perturbation enters during matter domination, the value of the decaying mode is zero, whereas if it enters during radiation domination, the decaying mode will quickly disappear. This reasoning cannot be applied directly in the setup adopted in this work. Indeed, we do not model the ``full'' $\Lambda$CDM (or $w_0w_a$CDM) framework\footnote{In this context, we refer to the ``full'' $\Lambda$CDM framework as the model in which the evolution of perturbations is followed from inflation all the way to the present time.
} starting from inflationary initial conditions.  Rather, our analysis focuses on the evolution of matter perturbations from $z \sim 1100$ onwards. Thus, we can adopt the results of the ``full'' model starting from that redshift—specifically, the fact that the decaying mode can be neglected—as a physically motivated justification for our assumption. 
We stress that the fact that we numerically start the integration at much earlier times is not in contrast with this reasoning.

We now assess the impact of including a decaying mode at $x_{\rm in}$. The physical predictions discussed in section~\ref{Sec:void_evolution} concern the evolution of cosmic structures at redshifts $z \lesssim 1100$, where the model is expected to accurately describe structure formation. The behavior of $\delta_{\rm L}(t)$ and $\delta_{\rm E}(t)$ at earlier times ($z \gg 1100$) is not \textit{physically} relevant, as it lies outside the regime where the model is intended to be applied. Consequently, it is possible to switch on a decaying mode at the initial time, as long as it becomes negligible by $z \sim 1100$ and the growing mode matches that of the standard $B = 0$ case. This requirement is essential. Thus, not all initial conditions with a decaying component are acceptable. Only those configurations in which the decaying mode dies off sufficiently early and leaves the correct growing mode in place lead to a physically meaningful evolution. In such cases, the subsequent dynamics are entirely governed by the growing mode and are indistinguishable from those obtained without any decaying mode at initial time.

The possibility of switching on a decaying mode at the initial time is justified by the fact that, during matter domination, the decaying mode scales as $a^{-3/2}(t)$, while the growing mode increases as $a(t)$ (see eq.~\eqref{Eq:matter_dominance_solution}). As a result, we expect that the earlier the initial conditions are set, the wider the range of decaying-mode amplitudes that still lead to the same physical outcome at lower redshift. We will make this statement more precise below.

The invariance of the physical results holds as long as two conditions are met: (i) the system is in the linear regime at $x_{\rm in}$, and (ii) the coefficient $A$ of the growing mode remains negative. When these conditions are satisfied, setting initial conditions further in the past increases the flexibility without affecting the physical outcome. 
Strictly speaking, however, the two conditions are not logically independent. In particular, violating condition (ii), i.e.~$A\geq0$, while keeping a negative initial density contrast for the void ($\delta_{\rm v,in}<0$) implies that the system is no longer in the linear regime. Indeed, if the evolution were still linear, the decaying mode would naturally fade away, and the solution would be dominated by the growing mode. A positive growing-mode coefficient would then inevitably lead to a sign switch in the density contrast at late times, which is completely unphysical. This inconsistency signals that the system at $x_{\rm in}$ is no longer described accurately by linear dynamics, and thus the standard linear interpretation cannot be applied.

One way to allow for a non-zero decaying mode at initial time is by adopting the following parametrization:
\begin{align}
    \begin{cases}
        \delta_{\rm E}(x_{\rm in}) &= \delta_{\rm v,in} \\
        \delta^{'}_{\rm E}(x_{\rm in}) &= \delta_{\rm v,in}\,p
    \end{cases}\,.
\label{Eq:parametrization_ICs_p}
\end{align}
where $p>-1.5$ is a real number that, if equal to unity, reduces the initial conditions to those of eq.~\eqref{Eq:the_initial_conditions_delta}. 
Solving for $A$ and $B$ as a function of $p$ and $\delta_{\rm v,in}$ we have
\begin{align}
    A \,=\, \frac{3}{5}\,\delta_{\rm v,in}\left(1 + \frac{2}{3}p\right)\,, \qquad B \,=\, \frac{2}{5}\,\delta_{\rm v,in}\left(1 - p\right)\,,
\end{align}
from which we understand that if $p\leq-1.5$ and $\delta_{\rm v,in}<0$, the coefficient of the growing mode $A$ becomes positive. 

The choice of parametrization adopted in eq.~\eqref{Eq:parametrization_ICs_p} is by no means unique. One could equally adopt alternative relations between $\delta_{\rm E}(x_{\rm in})$ and $\delta_{\rm E}^{'}(x_{\rm in})$, or even treat both quantities as fully independent. In that case, the initial conditions would be determined via a two-dimensional shooting to match the desired late-time value of $\delta_{\rm E}$.
The main advantage of the parametrization in eq.~\eqref{Eq:parametrization_ICs_p} is practical: once a value of $p$ is fixed, the shooting becomes one-dimensional and can be implemented in the numerical code in complete analogy with the standard case. 
While the parametrization itself does not have a direct physical meaning, the parameter $p$ can be interpreted in terms of the ratio between the amplitudes of the growing and decaying modes, defined as
\begin{align}
    r \,\equiv\, \frac{B}{A} \,=\, \frac{2(1-p)}{3(1+\tfrac{2}{3}p)} \,.
    \label{Eq:ratio_r}
\end{align}
\begin{figure}[t]
    \centering
    \includegraphics[width=0.95
    \linewidth]{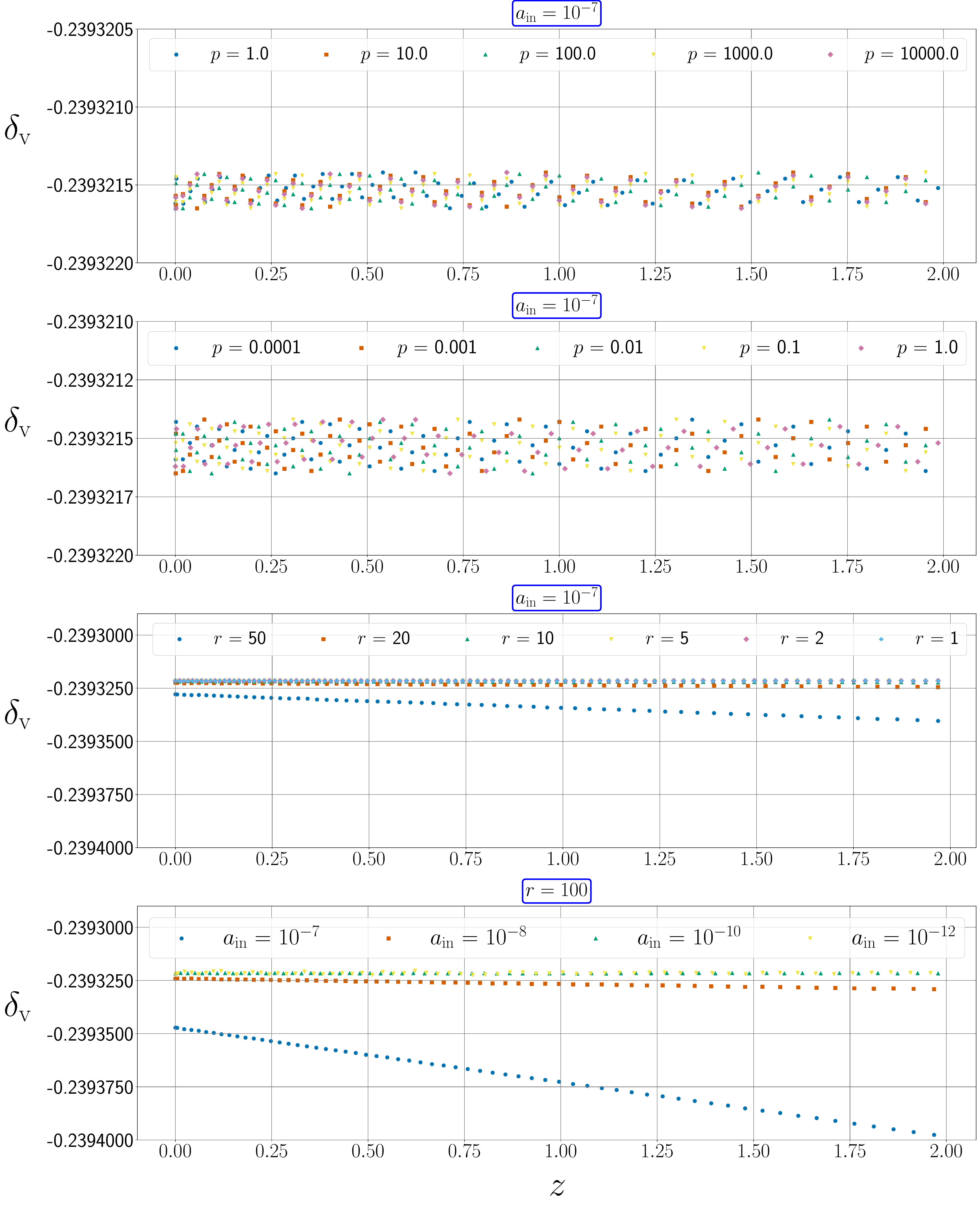}
    \caption{Redshift dependence of the mapping $\delta_{\rm v,in}(z, \delta_{\rm E} = -0.2)$ in the range $z \in [0,1]$. The figure is organized in four rows: the first two vary the parameter $p$ for fixed $a_{\rm in} = 10^{-7}$, with $p = 1$ to $10^4$ (first row) and $p = 1$ to $10^{-4}$ (second row); the third row shows variation in terms of $r = 1\,, 2\,, 5\,, 20\,, 50\,$; the fourth row fixes $r = 100$ and varies the starting time as $a_{\rm in} = 10^{-7}\,, 10^{-8}\,, 10^{-10}\,, 10^{-12}$.}
    \label{Fig:test_p}
\end{figure}
We now investigate how varying the parameter $p$ in eq.~\eqref{Eq:parametrization_ICs_p} affects the evolution of the system.
To this end, we study the Lagrangian mapping $\delta_{\rm v}(z,\delta_{\rm E})$ for fixed $\delta_{\rm E} = -0.2$ in the redshift interval $z \in [0,1]$, considering an EdS universe. The choice $\delta_{\rm E} = -0.2$ is not physically motivated; any value above the shell-crossing threshold would be equally suitable. The results are shown in figure~\ref{Fig:test_p}.
We consider five representative regimes for $p$.

\paragraph{Case I: $p < -1.5$} Since, in the numerical code, we consider initial conditions with $\delta_{\rm v,in} < 0$, this configuration corresponds to having a positive growing mode ($A > 0$) and a negative decaying mode ($B < 0$). In this case, however, the assumption of linearity cannot be valid. Thus, the early-time solution can no longer be described by eq.~\eqref{Eq:matter_dominance_solution}. Indeed, if the system were truly in the linear regime (in which we can describe the dynamics through eq.~\eqref{Eq:matter_dominance_solution}), the growing mode would dominate and lead to a positive density contrast, i.e., an overdensity. This is clearly inconsistent with our choice of initial conditions ($\delta_{\rm v,in}<0$) and with the physical interpretation of a void. For this reason, we exclude this range of $p$ values from the analysis.

\paragraph{Case II: $p = -1.5$}

In this case, if the system were in the linear regime, the solution would contain no growing mode at all. As a result, the perturbation would simply decay over time, and the entire discussion about setting initial conditions to reproduce a specific late-time configuration becomes irrelevant. Without a growing mode, there is no meaningful way to describe the formation or evolution of a void within our framework.

\paragraph{Case III: $p > 1$}
This regime corresponds to a situation in which the growing mode is negative ($A<0$) and the decaying mode is positive ($B>0$). As $p$ increases, the ratio $r\rightarrow-1$ implies a fixed relative contribution between the two modes. For large values of $p$, this configuration leads to very large initial derivatives, but the results of the code remain fully stable. We find that the final results are insensitive to this choice: the difference in $\delta_{\rm v}$ remains below $10^{-5}\%$ even for extreme values such as $p = 10^4$. Physically, this means that amplifying the derivative while preserving the same initial density leads to the same evolution. This is illustrated in the first row of figure~\ref{Fig:test_p}, where we vary $p = 1, 10, 10^2, 10^3, 10^4$.

\paragraph{Case IV:} $0 < p < 1$
In this case, both $A$ and $B$ are negative, implying that we are activating a decaying mode of the same sign as the growing mode. As $p$ decreases, the ratio $r$ approaches a constant value $r \rightarrow 2/3$. This saturation means that pushing $p$ to smaller and smaller values does not introduce any new physical effects. In practice, this scenario corresponds to setting the initial derivative to a value much smaller than the density contrast. Our results confirm that in this regime, the solution becomes insensitive to the exact value of the derivative. This is shown in the second row of figure~\ref{Fig:test_p}, where we test values $p = 10^{-4}, 10^{-3}, 10^{-2}, 10^{-1}, 1$. 

\paragraph{Case V:} $-1.5 < p < 0$
Both $A$ and $B$ are negative, but as $p \rightarrow -1.5$, $r$ can become arbitrarily large and the decaying mode dominates over the growing one at $x_{\rm in}$.
In the third row of figure~\ref{Fig:test_p}, we plot the results for $\delta_{\rm v}$ varying the value of $r$ (or $p$ equivalently) starting the integration at $x_{\rm in}.$ We find that for small values of $r$ (of order unity), the results are identical to the standard case. However, as $r$ increases beyond that, the solution starts to show a linear dependence at the fifth decimal place. While these are completely irrelevant for any practical application, we find it worthwhile to investigate their theoretical origin.

What we observe is that for large $r$, the mapping $\delta_{\rm v}(z)$ acquires a weak linear dependence on redshift, breaking the expected constancy in the EdS case. To further understand this effect, we perform an additional test: we fix $r = 100$ and repeat the integration starting from increasingly earlier times. As shown in the fourth row of figure~\ref{Fig:test_p}, the solutions converge as we move the starting point further back in time. This confirms that the impact of a large decaying mode at early times is suppressed by the time evolution, and becomes irrelevant for the redshift range of interest.

In conclusion, our results confirm that for any choice of $p > -1.5$, the final value of $\delta_{\rm v}(z)$ in the redshift range $z \in [0,1]$ remains unchanged, provided that the integration is started sufficiently early. This validates the qualitative statement made at the beginning: the earlier the initial time is set, the more irrelevant the presence of a decaying mode becomes. This behaviour is clearly observed in the plots and provides a direct justification for treating the decaying mode as negligible in our analysis. Although the solutions differ near the initial integration point, these differences do not have any physically relevant impact for $z\lesssim1100$. For the purposes of the present work, this confirms that setting the decaying mode to zero is fully justified, as its presence or absence becomes physically irrelevant under the conditions outlined above.

\section{Consistency checks with the EdS model}
\label{Sec:Consistency_checks_with_the_EdS_model}
In this appendix, we apply our numerical solver to the EdS model in order to perform a series of consistency checks.  These serve as benchmarks to test the accuracy and validity of the code and to ensure that the results agree with the theoretical predictions. We use the EdS model because it admits fully analytical solutions, which allow us to make a direct and reliable comparison with our numerical results. These tests are particularly meaningful because, when the analysis is extended to other cosmological scenarios, i.e.~$\Lambda$CDM and $w_0w_a$CDM, the only difference (in the code) lies in the background evolution of the matter density parameter $\Omega_\mathrm{m}(z)$ in eq.~\eqref{Eq:non_linear_evolution_equation}. We present the results of the following checks.
\paragraph{First check.} The two methods, that is, the  $R$-based and the hydrodynamical one, yield identical results, and both match the analytical solution when studying the evolution of an isolated void. 
\begin{figure}[t]
    \centering
    \includegraphics[width=\textwidth]{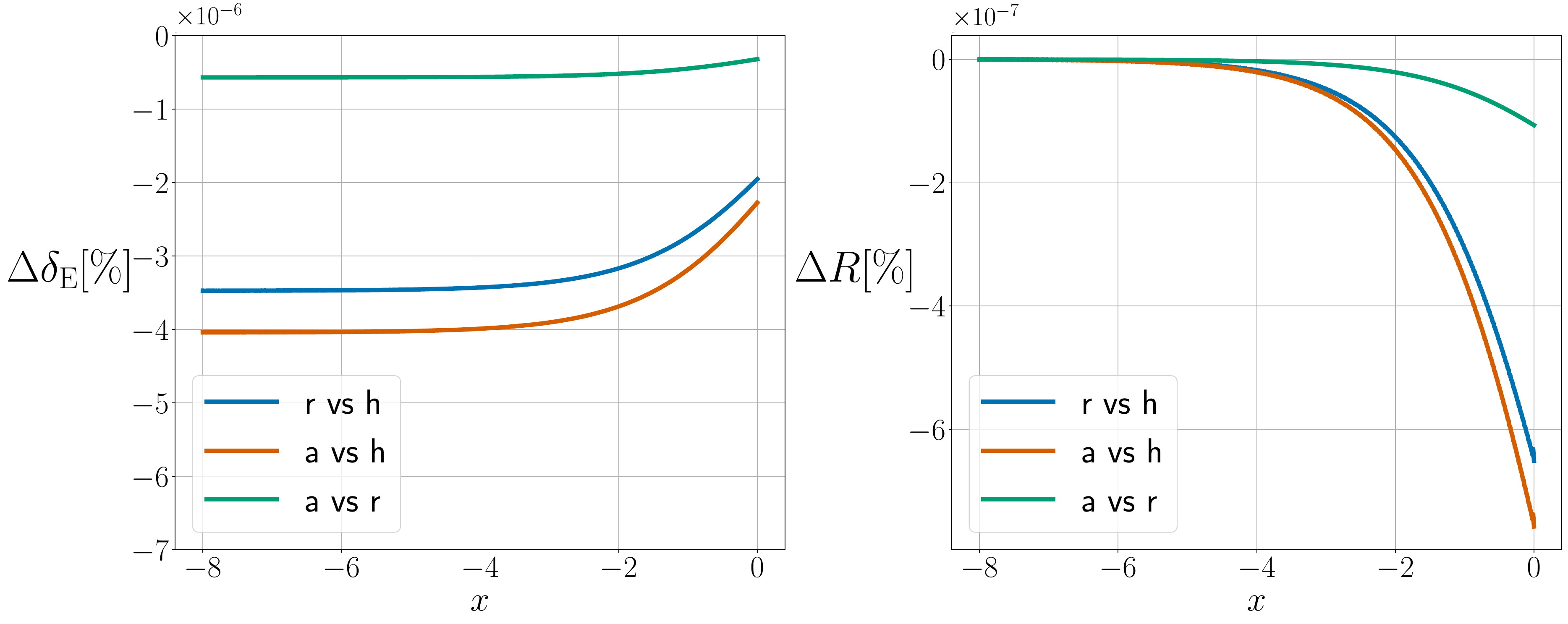}
        \caption{Left panel: we show the percentage differences in the solutions for $\delta_{\rm E}$, denoted as $\Delta\delta_{\rm E}[\%]$, obtained using the analytical (a) method, and the numerical $R$-based (r) and hydrodynamical (h) methods. All pairwise comparisons are shown (i.e., a vs r, a vs h, and r vs h) for an isolated void that reaches $\delta_{\rm E} = -0.5$ at $z = 0$.
        Right panel: same as in the left panel, but for the radius $R$, with percentage differences denoted as $\Delta R[\%]$.
        }
        \label{Fig:delta_check_d=-0.5}
\end{figure}

In figure~\ref{Fig:delta_check_d=-0.5} we study the case of a void reaching $\delta_{\rm E} = -0.5$ today ($z=0$). We present the solution starting from $x=-8\,$, that is, $z\approx 3000$ up to present time ($x=0$). This choice allows us to plot the solutions in the redshift range in which we are interested. We compare the solutions obtained analytically (a, for analytical) with those derived numerically using either the $R$-based (r, for $R$-based) or hydrodynamical (h, for hydrodynamical) approach. 
We present the relative percentage differences for both $\delta_{\rm E}$ and $R$, denoted as $\Delta\delta_{\rm E}[\%]$ and $\Delta R[\%]$, computed by comparing the results obtained with the three different methods (a, r, and h). For example, when comparing the solutions obtained with the analytical (a) and hydrodynamical (h) methods, the legend in the plot indicates ``a vs h'', and the corresponding differences are computed as
\begin{align}
    \Delta\delta_{\rm E}[\%] \equiv \frac{\delta_{\rm E}^{\rm (a)} - \delta_{\rm E}^{\rm (h)}}{\delta_{\rm E}^{\rm (h)}} \times 100\,,
\end{align}
where the superscript indicates the method used to compute the solution.

These quantities quantify how much the predicted evolution of the density contrast $\delta_{\rm E}$ and radius $R$ differ between methods.
As shown, the solutions for $\delta_{\rm E}$ and $R$ differ by less than $(10^{-6})\%$, confirming that three methods are fully consistent in an EdS universe. Such precision is irrelevant for applications involving real data or simulations, but it serves to demonstrate the theoretical equivalence of the approaches. The behaviour observed in the plots does not appear to be purely numerical: the percentage difference shows a dependence on $x$, which likely results from the accumulation of numerical errors during integration. However, this effect emerges at precision levels far beyond what is achievable in realistic scenarios, well beyond the accuracy of the results presented in this work, and is therefore not discussed further.
\paragraph{Second Check.} The second check we present is whether the two shell-crossing conditions, i.e.~eqs.~\eqref{Eq:Shell_Crossing_R_R_right_hand_derivative} and \eqref{Eq:Shell_crossing_Delta}, we have derived in section~\ref{Sec:Shell_Crossing} are consistent with each other and with the analytic result in eq.~\eqref{Eq:EdS_shell_crossing_value}, i.e.~$\delta_{\rm v}(z,\delta_{\rm E,sc}) = -2.71718$. We recall that this value is redshift-independent in an EdS universe.
\begin{figure}[t!]
    \centering
    \includegraphics[width=1.0\textwidth]{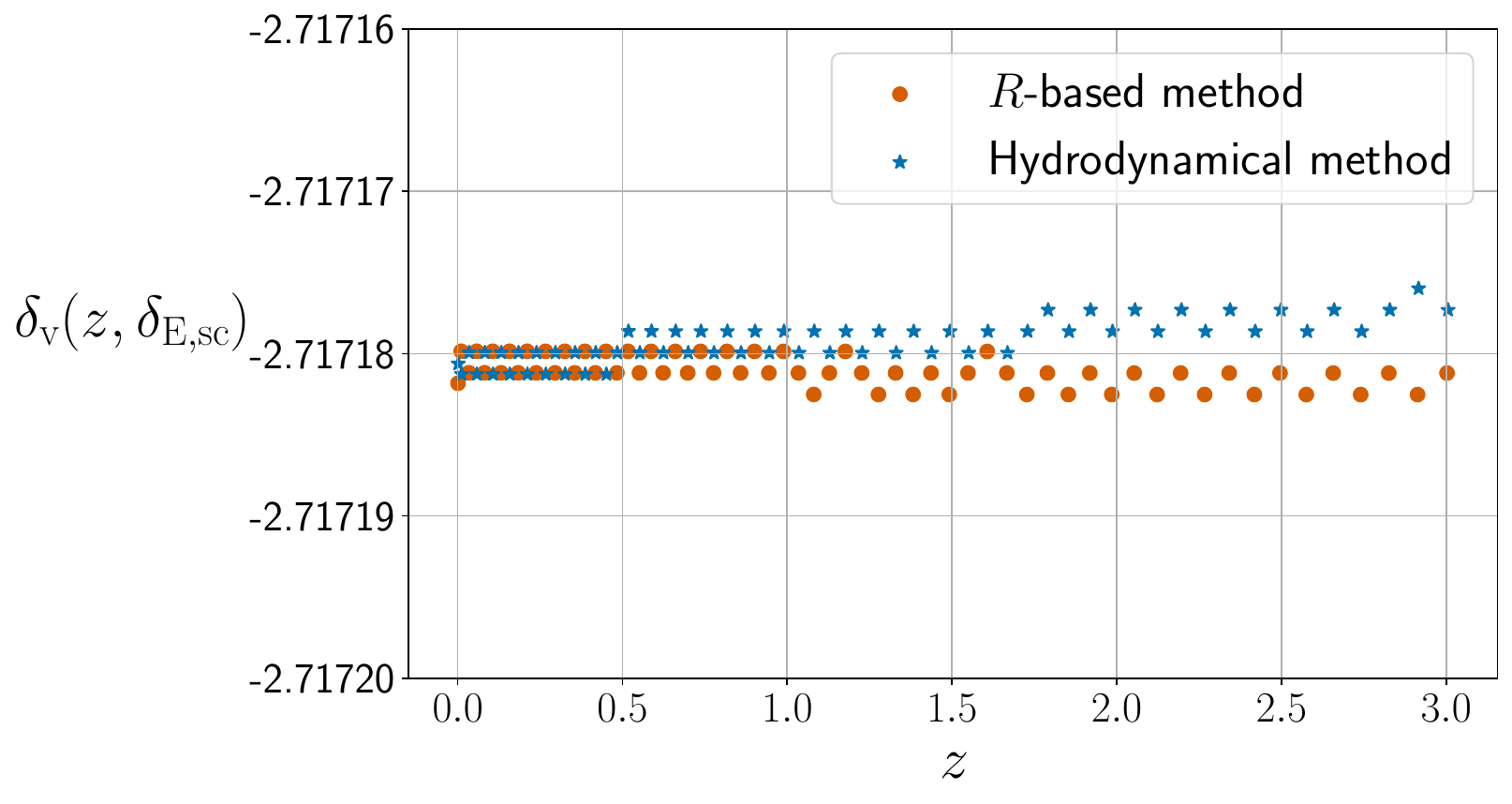}
    \caption{We present the linearly extrapolated matter density perturbation at the moment of shell-crossing ($\delta_{\rm E} = \delta_{\rm E,sc})$, i.e.~$\delta_{\rm v}(z,\delta_{\rm E,sc})$, computed using both the $R$-based and hydrodynamical methods, for voids that undergo shell-crossing in the redshift range $z \in (0, 3)$ in an EdS universe.}
    \label{Fig:comparison_shell_crossing_conditions}
\end{figure}

Thus, we examine the evolution of the linearly extrapolated matter density perturbation $\delta_{\rm v}(z, \delta_{\rm E,sc})$ as a function of redshift $z$, for voids making shell-crossing in $z \in (0, 3)$. The plots are obtained using the same procedure described in section~\ref{Sec:implementation_of_the_shell_crossing}. The only difference is that, when using the $R$-based method, we solve eq.~\eqref{Eq:Newton_equation} up to the shell-crossing time computed via eq.~\eqref{Eq:Shell_Crossing_R_R_right_hand_derivative}, while the linear value is always extracted using eq.~\eqref{Eq:linear_evolution_equation}. 
The results presented in figure~\ref{Fig:comparison_shell_crossing_conditions} demonstrate a remarkable agreement among the different methods, confirming the validity of the two derived shell-crossing conditions. 

\begin{figure}[t!]
    \centering
    \includegraphics[width=1.0\linewidth]{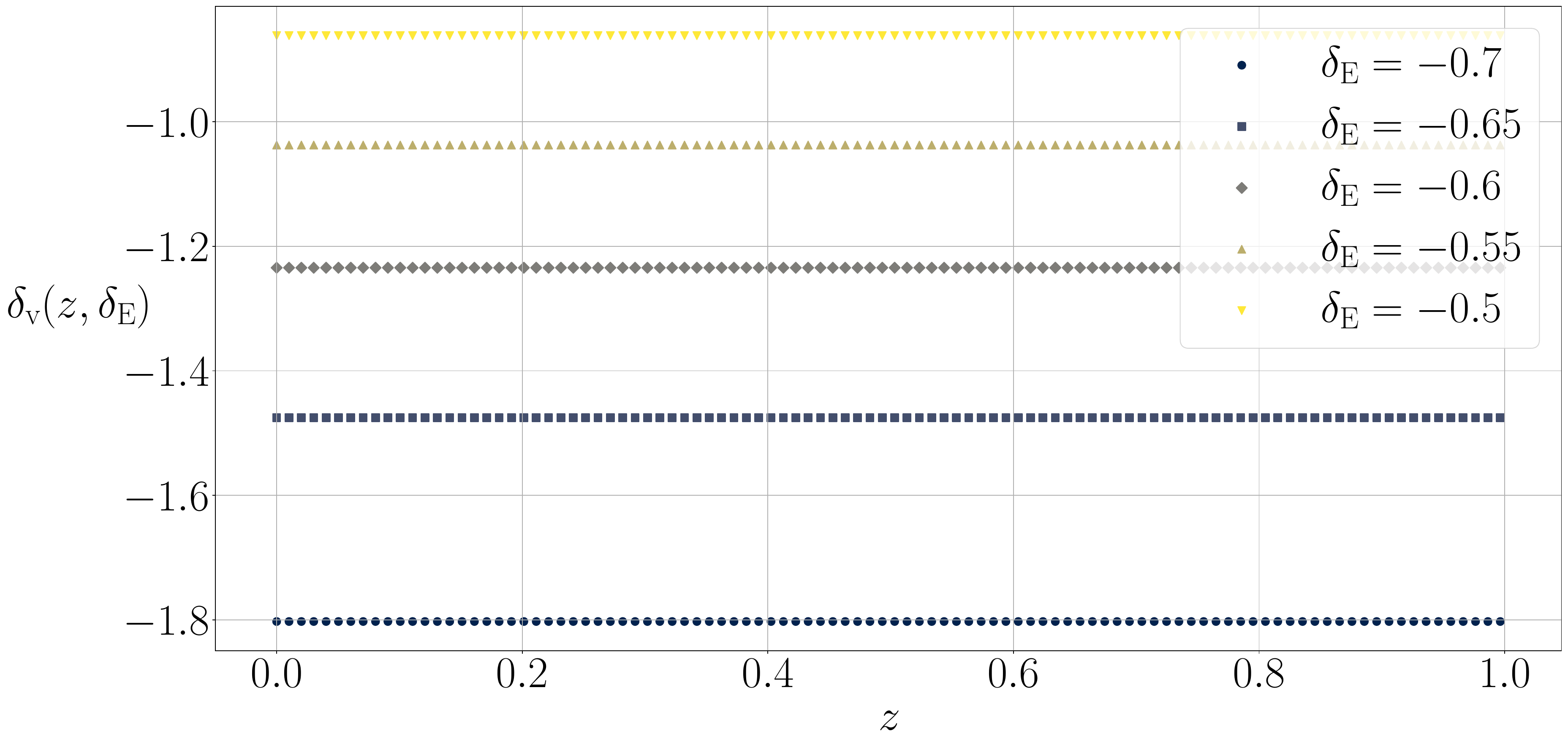}
    \caption{We present the redshift dependence of the linear to non-linear mapping $\delta_{\rm v}(z,\delta_{\rm E})$ in an EdS universe, keeping $\delta_{\rm E}$ fixed to selected values, i.e.~$\delta_{\rm E} = [-0.7,\,-0.65,\,-0.6,\,-0.55,\,-0.5]\,$. These results were obtained using the hydrodynamical formalism introduced in section~\ref{Sec:hydrodynamical_approach}.}
    \label{fig:map_EdS_constant}
\end{figure}

\paragraph{Third Check. } The final check we present concerns the fact that the linear to non-linear mapping, i.e., $\delta_{\rm v}(z, \delta_{\rm E})$, discussed in section~\ref{Sec:the_linear_to_non_linear_mapping}, is redshift-independent in an EdS universe. This is a well-known result, which our numerical solver reproduces with remarkable accuracy, as shown in figure~\ref{fig:map_EdS_constant}. The results in the plot are obtained using the hydrodynamical formalism.

\bibliographystyle{JHEP}
\bibliography{main.bib}

\end{document}